\documentclass[a4paper,11pt]{article}

\usepackage{graphicx}
\usepackage{color}
\usepackage{version}
\usepackage[colorlinks]{hyperref}
\usepackage[normalem]{ulem}

\usepackage{lcbmaths}
\usepackage{wasysym}

\DeclareMathOperator{\Covop}    {\Sigma}
\newcommand{\Cov}[2]            {\Covop\!\bracr{{#1},{#2}}}
\newcommand{\Covs}[1]           {\Covop\!\bracr{#1}}
\newcommand{\Covc}[2]           {\Covop\!\bracr{\lcond{#1}{#2}}}

\DeclareMathOperator{\trop}     {tr}
\renewcommand{\trace}[1]        {\trop\!\bracr{#1}}
\renewcommand{\dett}[1]         {\bracp{#1}}

\newcommand{\bX}                {\boldsymbol X}
\newcommand{\bY}                {\boldsymbol Y}

\newcommand{\bZ}                {\boldsymbol Z}

\newcommand{\btZ}               {\tilde\bZ}
\newcommand{\blX}               {\bX^{\!-}}
\newcommand{\blY}               {\bY^{\!-}}
\newcommand{\blZ}               {\bZ^{\!-}}
\newcommand{\bltZ}              {\btZ^{\!-}}
\newcommand{\bU}                {\boldsymbol U}
\newcommand{\bV}                {\boldsymbol V}
\newcommand{\bW}                {\boldsymbol W}

\newcommand{\bx}                {\boldsymbol x}
\newcommand{\by}                {\boldsymbol y}

\newcommand{\eps}               {\varepsilon}
\newcommand{\beps}              {\boldsymbol\eps}
\newcommand{\beeta}             {\boldsymbol\eta}
\newcommand{\bxi}               {\bm\xi}
\newcommand{\mc}                {\oplus}

\DeclareMathOperator{\TE}       {\mathcal T}
\DeclareMathOperator{\GC}       {\mathcal F}

\DeclareMathOperator{\EGC}      {\widehat\GC}

\newcommand{\TGC}               {\GC^{tr}}
\newcommand{\PGC}               {\GC^P}

\newcommand{\gc}[2]             {\GC_{{#1}\to{#2}}}
\newcommand{\cgc}[3]            {\GC_{\lcond{{#1}\to{#2}}{#3}}}
\newcommand{\ecgc}[3]           {\EGC_{\lcond{{#1}\to{#2}}{#3}}}
\newcommand{\tgc}[2]            {\TGC_{{#1}\to{#2}}}
\newcommand{\ctgc}[3]           {\TGC_{\lcond{{#1}\to{#2}}{#3}}}

\newcommand{\cpgc}[3]           {\PGC_{\lcond{{#1}\to{#2}}{#3}}}

\newcommand{\cxe}[3]            {\TE_{\lcond{{#1}\to{#2}}{#3}}}

\newcommand{\Secref}[1]         {Section~\ref{#1}}

\newcommand{\Apxref}[1]         {Appendix~\ref{#1}}

\definecolor{atncol}            {rgb}{0.6,0,0}

\newcommand{\SGC}               {f}
\newcommand{\sgc}[2]            {\SGC_{{#1}\to{#2}}}

\newcommand{\TSGC}              {f^{tr}}
\newcommand{\tsgc}[2]           {\TSGC_{{#1}\to{#2}}}

\newcommand{\PSGC}              {\SGC^P}

\newcommand{\pscgc}[3]          {\PSGC_{\lcond{{#1}\to{#2}}{#3}}}
\newcommand{\dlam}              {\,\dop\!\lambda}

\DeclareMathOperator{\CD}       {cd}
\DeclareMathOperator{\BCD}      {bcd}
\DeclareMathOperator{\TCD}      {tcd}

\newcommand{\cd}[1]				{\CD\!\bracr{#1}}
\newcommand{\cdx}[2]			{\CD_{#1}\!\bracr{#2}}
\newcommand{\bcd}[1]			{\BCD\!\bracr{#1}}

\newcommand{\tcd}[1]			{\TCD\!\bracr{#1}}

\begin{document}

\author{Adam B.~Barrett$^1$\footnote{email: Adam.Barrett@sussex.ac.uk},
 Lionel Barnett$^2$\footnote{email: L.C.Barnett@sussex.ac.uk}
  \setcounter{footnote}{3}
  and Anil K.~Seth$^1$\footnote{email: A.K.Seth@sussex.ac.uk} \\
  \\
    $^1$\textit{Sackler Centre for Consciousness Science, School of Informatics,}\\
   University of Sussex,
    Brighton BN1 9QJ, UK\\
\\
   $^2$\textit{Centre for Computational Neuroscience and Robotics,
    School of Informatics,}\\
   University of Sussex,
    Brighton BN1 9QJ, UK}

\title{Multivariate Granger Causality and Generalized Variance}

\maketitle

\begin{abstract}
\noindent Granger causality analysis is a popular method for
inference on directed interactions in complex systems of many
variables. A shortcoming of the standard framework for Granger
causality is that it only allows for examination of interactions
between single (univariate) variables within a system, perhaps
conditioned on other variables. However, interactions do not
necessarily take place between single variables, but may occur
among groups, or ``ensembles'', of variables. In this study we
establish a principled framework for Granger causality in the
context of causal interactions among two or more multivariate
sets of variables. Building on Geweke's seminal 1982 work, we offer
new justifications for one particular form of multivariate Granger
causality based on the generalized variances of residual errors. Taken together, our results support a comprehensive and theoretically consistent extension of Granger causality to the multivariate case.  Treated individually, they highlight several specific advantages of the generalized variance measure, which we illustrate using applications in neuroscience as an example. We further show how the measure can be used to define ``partial''
Granger causality in the multivariate context and we also motivate
reformulations of ``causal density'' and ``Granger autonomy''. Our
results are directly applicable to experimental data and promise to
reveal new types of functional relations in complex systems, neural
and otherwise.
\\

\\


\end{abstract}

\section{Introduction}

A key challenge across many domains of
science and engineering is to understand the behavior of complex
systems in terms of dynamical interactions among their component
parts. A common way to address this challenge is by analysis of time
series data acquired simultaneously from multiple system components.
Increasingly, such analysis aims to draw inferences about
\emph{causal} interactions among system variables
\cite{DingEtal06,Friston03,Schreiber00}, as a complement to standard
assessments of undirected functional connectivity as revealed by
coherence, correlation, and the like.

A first step in any dynamical analysis is to identify target
variables. Typically, subsequent analysis then assumes that
functional (causal) interactions take place among these variables.
However, in the general case it may be that explanatorily relevant
causal interactions take place among \emph{groups}, or ``ensembles'',
of variables \cite{ladroue:2009,Geweke82}. It is important to
account for this possibility for at least two reasons.  First,
identification of target variables is usually based on \emph{a
priori} system knowledge or technical constraints, which may be
incomplete or arbitrary, respectively. Second, even given
appropriate target variables, it is possible that relevant
interactions may operate at multiple scales within a system, with
larger scales involving groups of variables.   Consider an example
from functional neuroimaging.  In a typical fMRI\footnote{Functional
magnetic resonance imaging.} study, the researcher may identify
\emph{a priori} several ``regions-of-interest'' (ROI) in the brain,
each represented in the fMRI dataset by multiple voxels, where each
voxel is a variable comprising a single time series reflecting
changes in the underlying metabolic signal.  Assuming that the
objective of the study is to assess the causal connectivity among
the ROIs, a standard approach is to derive a single time series for
each ROI either by averaging or by extracting a principal component
\cite{zhou:2009}; alternatively, repeated pairwise analysis can be
performed on each pair of voxels.  A more appropriate approach,
however, may be to consider causal interactions among the
multivariate groups of voxels comprising each ROI.  Similar
scenarios could be concocted in a very wide range of application
areas, including economics, biology, climate science, among others.

In this paper, we describe a principled approach to assessing causal
interactions among multivariate groups of variables. Our approach is
based on the concept of Granger causality (G-causality)
\cite{wiener:1956,Granger69}, a statistical notion of causality
which originated in econometrics but which has since found
widespread application in many fields, with a particular
concentration in the neurosciences \cite{DingEtal06,bressler:2010}.
G-causality is an example of time series inference on stochastic
processes and is usually implemented via autoregressive modeling of
multivariate time series.  The basic idea is simple: one variable
(or time series) can be called ``causal'' to another if the ability to
predict the second variable is improved by incorporating information
about the first.  More precisely, given inter-dependent variables
$X$ and $Y$, it is said that ``$Y$ G-causes $X$'' if, in a
statistically suitable manner, $Y$ assists in predicting the future
of $X$ beyond the degree to which $X$ already predicts its own
future.  It is straightforward to extend G-causality to the
conditional case \cite{Geweke82}, where $Y$ is said to G-cause $X$,
conditional on $Z$, if $Y$ assists in predicting the future of $X$
beyond the degree to which $X$ and $Z$ together already predict the
future of $X$.  Importantly, conditional G-causality is orthogonal
to the notion of inferring causality among groups of variables,
which is the focus of the present paper and which we term
\emph{multivariate} G-causality.  In the multivariate case, the
above description of G-causality is generalized to interactions
among \emph{sets} of interdependent variables $\bX$, $\bY$, and (in
the conditional multivariate case) $\bZ$.  The generalization we
propose was originally introduced in the field of econometrics by
Geweke in 1982 \cite{Geweke82}, but has since been almost totally
overlooked. Indeed a different measure has recently appeared
\cite{ladroue:2009}. In the following, we derive several
justifications for preferring Geweke's measure, some of which we examine numerically. We go on to explore a
series of implications for the analysis of complex systems in
general, with a particular focus on applications in neuroscience.

After laying out our conventions in Section 2, in Section 3 we
introduce two alternative measures of multivariate G-causality. The
formulations differ according to their treatment of the covariance
matrices of residuals in the underlying autoregressive models:
Geweke's measure uses the \textit{determinant} of this matrix (the
generalized variance), while the other uses the \textit{trace} (the
total variance). Section 4 explores several advantageous properties
of the determinant formulation as compared to the trace formulation.
In brief, the determinant formulation is fully equivalent to
transfer entropy \cite{Schreiber00} under Gaussian assumptions, is
invariant under a wider range of variable transformations, is
expandable as a sum of standard univariate G-causalities, and admits
a satisfactory spectral decomposition. Numerically, we show that Geweke's measure is just as stable as is the alternative measure based on the total variance. Section 5 extends the
determinant formulation to the important case of ``partial''
G-causality which provides some measure of control with respect to
unmeasured latent or exogenous variables.  Section 6 extends a
previously defined measure of ``causal density'' \cite{seth:2009a}
which reflects the overall dynamical complexity of causal
interactions sustained by a system.  In Section 7 we show how
multivariate G-causality can enhance a measure of ``autonomy'' (or
``self-causation'') based on G-causality \cite{seth:alife:2009}, and
Section 8 carries the discussion towards the identification of
macroscopic variables via the notion of causal independence. Section
9 provides a general discussion and summary of contributions.

\section{Notational conventions and preliminaries}

We use a mathematical vector/matrix notation in which bold type generally denotes vector quantities and upper-case type denotes matrices or random variables, according to context. All vectors are considered to be \textit{column} vectors. `$\mc$' denotes \textit{vertical concatenation} of vectors, so that for $\bx = \trans{(x_1,\ldots,x_n)}$ and $\by = \trans{(y_1,\ldots,y_m)}$, $\bx\mc\by$ is the vector $\trans{(x_1,\ldots,x_n,y_1,\ldots,y_m)}$ of dimension  $n+m$, where the symbol `$\transop$' denotes the transpose operator. We also write $\dett\cdot$ for the determinant and $\trace\cdot$ for the trace of a square matrix.

Given jointly distributed multivariate random variables (\ie\ random vectors) $\bX,\bY$, we denote by $\Covs\bX$ the $n \times n$ matrix of covariances $\cov{X_i}{X_j}$ and by $\Cov\bX\bY$ the $n \times m$ matrix of cross-covariances $\cov{X_i}{Y_\alpha}$. We then use $\Covc\bX\bY$ to denote the $n \times n$ matrix
\begin{equation}
    \Covc\bX\bY \equiv \Covs\bX - \Cov\bX\bY \Covs\bY^{-1} \trans{\Cov\bX\bY}\,, \label{eq:ccxy}
\end{equation}
defined when $\Covs\bY$ is invertible. $\Covc\bX\bY$ appears as the covariance matrix of the residuals of a linear regression of $\bX$ on $\bY$ (c.f.~Eq.~\eqref{eq:rescov} below); thus, by analogy with \textit{partial correlation} \cite{Kendall79} we term $\Covc\bX\bY$ the \textit{partial covariance}\footnote{This is to be distinguished from the \textit{conditional covariance}, which will in general be a random variable, though later we note that for \textit{Gaussian} variables the notions coincide.} of $\bX$ given $\bY$. Similarly, given another jointly distributed variable $\bZ$, we define the \textit{partial cross-covariance}
\begin{equation}
    \Covc{\bX,\bY}\bZ \equiv \Cov\bX\bY - \Cov\bX\bZ \Covs\bZ^{-1} \trans{\Cov\bY\bZ}\,. \label{eq:cccxy}
\end{equation}

The following identity \cite{Barnett:2009a} will be useful for deriving certain properties of multivariate G-causality:
\begin{equation}
    \dett{\Covc\bX\bY} = \left. \dett{\Covs{\bX\mc\bY}} \right/ \dett{\Covs\bY}\,. \label{eq:cxyid}
\end{equation}

Suppose we have a multivariate stochastic process $\bX_t$ in discrete time\footnote{While our analysis may be extended to \textit{continuous} time we focus here on the discrete time case.} (\ie\ the random variables $X_{it}$ are jointly distributed). We use the notation $\bX^{(p)}_t \equiv \bX_t\mc\bX_{t-1}\mc\ldots\mc\bX_{t-p+1}$ to denote $\bX$ itself, along with $p-1$ \textit{lags}, so that for each $t$, $\bX^{(p)}_t$ is a random vector of dimension $pn$. Given the lag $p$, we also often use the shorthand notation $\blX_t \equiv \bX^{(p)}_{t-1}$ for the lagged variable.

\section{Multivariate Granger causality} \label{sec:granger}
G-causality analysis is concerned with the comparison of different linear regression models of data. Thus, let us consider the (multivariate) linear regression of one random vector $\bX$, the predictee, on another random vector $\bY$, the predictor:\footnote{Here and in the remainder of this paper we assume, without loss of generality, that all random vectors and random processes have zero mean; thus constant terms are omitted in all linear regressions.}
\begin{equation}
    \bX = A \cdot \bY + \beps\,, \label{eq:linreg}
\end{equation}
where the $n \times m$ matrix $A$ contains the regression coefficients and the random vector $\beps = \trans{(\eps_1,\ldots,\eps_n)}$ comprises the residuals. The coefficients of this model are uniquely specified by imposing zero correlation between the residuals $\beps$ and the regressors (predictors) $\bY$. Via the Yule-Walker procedure \cite{DingEtal06,Barnett:2009a} one obtains
\begin{equation}
    A =  \Cov\bX\bY \Covs\bY^{-1} \label{eq:regcoef}
\end{equation}
and finds the covariance matrix of the residuals to be given by
\begin{equation}
    \Covs\beps = \Covc\bX\bY\,, \label{eq:rescov}
\end{equation}
with $\Covc\bX\bY$ defined as in \eqref{eq:ccxy}.

Suppose now we have three jointly distributed,
stationary\footnote{The analysis carries through for the
non-stationary case, but for simplicity we assume here that all
processes are stationary.} multivariate stochastic processes $\bX_t,
\bY_t, \bZ_t$. Then to measure the G-causality from $\bY$ to
$\bX$ given $\bZ$, one wants to compare the following two
multivariate autoregressive (MVAR) models for the processes
\cite{Granger69}:
\begin{equation}
\begin{split}
    \bX_t & = A \cdot \bracr{\bX^{(p)}_{t-1}\mc\bZ^{(r)}_{t-1}}  + \beps_t\,, \\
    \bX_t & = A' \cdot \bracr{\bX^{(p)}_{t-1}\mc\bY^{(q)}_{t-1}\mc\bZ^{(r)}_{t-1}} + \beps'_t\,.
\end{split} \label{eq:reg}
\end{equation}
Thus the predictee variable $\bX$ is regressed firstly on the previous $p$ lags of itself plus $r$ lags of the conditioning variable $\bZ$ and secondly, in addition, on $q$ lags of the predictor variable $\bY$ (in theory, if not in practice, $p$, $q$ and $r$ could be infinite).\footnote{This might be more familiar as \textit{conditional} G-causality, with $\bZ$ the conditioning variable. In practice it is the more useful form; for the non-conditional version, $\bZ$ may simply be omitted.}

The standard measure of G-causality used in the literature is defined only for \textit{univariate} predictor and predictee variables $Y$ and $X$, and is given by the log of the ratio of the residual variances for the regressions \eqref{eq:reg}. In our notation,\footnote{Note that even though $X$ and $Y$ are univariate, the \textit{lagged} variables $\blX$ and $\blY$ will generally be multivariate (at least if $p,q>1$); hence they are written in bold type.}
\begin{align}
    \cgc Y X\bZ
    &\equiv \lnt{\frac{\var{\eps_t}}{\var{\eps'_t}}} \nonumber \\
    &= \lnt{\frac{\Covs{\eps_t}}{\Covs{\eps'_t}}} \nonumber \\
    &= \lnt{\frac{\Covc X{\blX\mc\blZ}}{\Covc X{\blX\mc\blY\mc\blZ}}}\,, \label{eq:gcu}
\end{align}
where the last equality follows from the general formula \eqref{eq:rescov}. By stationarity this expression does not depend on time $t$. Note that the residual variance of the first regression will always be larger than or equal to that of the second, so that $\cgc Y X\bZ \ge 0$ always.
As regards statistical inference, it is known that the corresponding maximum likelihood estimator\footnote{We remark that for significance testing of G-causality it is quite common to use the appropriate $F$-statistic for the regressions \eqref{eq:reg} rather than $\cgc Y X\bZ$ itself \cite{Granger69,Seth:2010}; the quantities are in any case related by a monotonic transformation.}  $\ecgc Y X\bZ$ will have (asymptotically for large samples) a $\chi^2$-distribution under the null hypothesis  $\cgc Y X\bZ = 0$ \cite{Granger63,Whittle53}, and a non-central  $\chi^2$-distribution under the alternative hypothesis $\cgc Y X\bZ > 0$ \cite{Geweke82,Wald43}.

We now consider the case where predictee and predictor
variables are no longer constrained to be univariate, i.e.~multivariate G-causality. For a multivariate predictor,
Eq.~\eqref{eq:gcu} above (with $Y$ replaced by the bold-type $\bY$)
is a valid and consistent formula for G-causality. However,
for the case of a multivariate predictee there is not yet a
standard definition for G-causality. One possibility is to
simply use the multivariate mean square error (i.e.~total variance,
or expected squared length of the multivariate residual), leading
to
\begin{align}
    \ctgc\bY\bX\bZ
	&\equiv \lnt{\frac{\trace{\Covs{\beps_t}}}{\trace{\Covs{\beps'_t}}}} \nonumber \\
    &=      \lnt{\frac{\trace{\Covc{\bX}{\blX\mc\blZ}}}{\trace{\Covc{\bX}{\blX\mc\blY\mc\blZ}}}}\,. \label{eq:gct}
\end{align}
We call this the trace version of multivariate G-causality
(trvMVGC). As recently noted by Ladroue
and colleagues \cite{ladroue:2009} trvMVGC appears to be a natural
extension of G-causality to the multivariate case because total
variance is a common choice for a measure of  goodness-of-fit or
prediction error for a multivariate regression. Moreover, the
measure is always non-negative, reduces to \eqref{eq:gcu} when the
predictee variable is univariate, and the regression matrix
coefficients that render the residuals uncorrelated with the
regressors also minimize the total variance (this is just the
``ordinary least squares'' procedure, minimizing mean square error).
Nonetheless, an alternative originally proposed by Geweke
\cite{Geweke82} uses instead the \textit{generalized variance}
$\dett{\Covs{\beps_t}}$, which quantifies the \textit{volume} in
which the residuals lie. This leads to the measure
\begin{align}
    \cgc\bY\bX\bZ
    &\equiv \lnt{\frac{\dett{\Covs{\beps_t}}}{\dett{\Covs{\beps'_t}}}} \nonumber \\
    &=      \lnt{\frac{\dett{\Covc{\bX}{\blX\mc\blZ}}}{\dett{\Covc{\bX}{\blX\mc\blY\mc\blZ}}}}\,. \label{eq:gc}
\end{align}
Like trvMVGC, this measure is always non-negative, reduces to
\eqref{eq:gcu} when the predictee variable is univariate, and is
consistent with the autoregressive approach inasmuch as the
Yule-Walker regression matrix coefficients minimize the generalized
variance, $\dett{\Covs\beps}$, as well as the total variance, (see
\Apxref{apx:mindet} for a proof). Geweke \cite{Geweke82} lists a
number of motivations for taking $\cgc\bY\bX\bZ$ as given in
Eq.~\eqref{eq:gc} as the natural extension of G-causality to
the multivariate case. These include: (i) that the generalized
variance version \eqref{eq:gc} is invariant under (linear)
transformation of variables (see \Secref{sec:invariance}); and (ii)
that the maximum likelihood estimator of this quantity,
$\ecgc\bY\bX\bZ$, is asymptotically $\chi^2$-distributed for large
samples. In the following section we further justify this choice.
Since we advocate the use of Geweke's measure \eqref{eq:gc} of
multivariate G-causality we abbreviate this simply as MVGC
henceforth.

As remarked previously, the expression \eqref{eq:gc} defines \emph{conditional} MVGC. Geweke \cite{Geweke84} gives the following intuitively appealing expression for $\cgc\bY\bX\bZ$ in terms of unconditional MVGCs:
\begin{equation}
    \cgc\bY\bX\bZ \equiv \gc{\bY\mc\bZ}\bX -  \gc\bZ\bX\,; \label{eq:cgc}
\end{equation}
that is, the extent to which $\bY$ and $\bZ$ together cause $\bX$ less the extent that $\bZ$ on its own causes $\bX$. Note that this identity also holds for trvMVGC.

\section{Properties of Multivariate Granger causality} \label{sec:Justifications}

In the following subsections we discuss some properties of MVGC and further motivate Geweke's definition of this measure.

\subsection{Gaussian Equivalence with Transfer Entropy} \label{sec:gauss}
When all variables are Gaussian distributed, the MVGC $\cgc\bY\bX\bZ$ is fully equivalent to the transfer entropy $\cxe\bY\bX\bZ$, an information-theoretic notion of causality \cite{Barnett:2009a}, with a simple factor of 2 relating the two quantities,
\begin{equation}
    \cgc\bY\bX\bZ = 2\cxe\bY\bX\bZ\,. \label{eq:tegceq}
\end{equation}
Transfer entropy \cite{Schreiber00,KaiserSchreiber02} is defined by
the difference in entropies
\begin{equation}
    \cxe\bY\bX\bZ \equiv \centro{\bX}{\blX\mc\blZ} - \centro{\bX}{\blX\mc\blY\mc\blZ}\,, \label{eq:te}
\end{equation}
and quantifies the degree to which knowledge of the past of $\bY$
reduces uncertainty in the future of $\bX$. The equivalence \eqref{eq:tegceq} stems from the entropy of a Gaussian distribution being directly proportional to the logarithm of the determinant of its covariance matrix; and, furthermore, from any conditional entropy involving Gaussian variables being directly proportional to the logarithm of the determinant of the appropriate corresponding \emph{partial} covariance matrix (see \cite{Barnett:2009a} for details). Due to the use of the determinant being crucial for this relationship, for trvMVGC the equivalence holds only in the more restricted situation when the predictee variable is univariate.

In addition to motivating MVGC over trvMVGC, the equivalence
\eqref{eq:tegceq} also provides a justification for the use of
linear regression models in measuring causality. Transfer
entropy is naturally sensitive to nonlinearities in the data, a property
which is rightly seen as desirable for measures of causality and
which has motivated the development of several nonlinear extensions
to standard G-causality \cite{ChenEtal04,Marinazzo08}.
However, when data are Gaussian, the two linear regressions capture
all of the entropy difference that defines transfer entropy,
which implies that non-linear extensions to G-causality
are of no additional utility. Indeed for two multivariate
Gaussian variables $\bX$ and $\bY$, the partial covariance
$\Covop(\bX|\bY)$, which is the same quantity as the residual
covariance under linear regression, can be simply thought of as the
conditional covariance of $\bX$ given $\bY$, because
$\mathrm{cov}(\bX|\bY=\by)=\Covop(\bX|\bY)$ for all $\by$. Hence, for
Gaussian data, linear regression accounts for all the dependence of
the regressee on the regressor.

To demonstrate formally that a stationary Gaussian AR process must be \textit{linear}, consider a general stationary multivariate Gaussian process $\bX_t$ satisfying
\begin{equation}
    \bX_t = f\!\bracr{\bX_{t-1}^{(p)}} + \beps_t\,,
\end{equation}
where $f(\cdot)$ is some sufficiently well-behaved, possibly nonlinear function and the $\beps_t$ are independent of $\bX_{t-s}$ for $s =1,2,\ldots$. For any $t$ then, $\beps_t = \bX_t - f\!\bracr{\bX_{t-1}^{(p)}}$ is independent of $\bX_{t-1}^{(p)}$, so that, in particular, for any value $\bxi$ taken by $\bX_{t-1}^{(p)}$, the conditional expectation
\begin{equation}
    \cexpect{\beps_t}{\bX_{t-1}^{(p)} = \bxi} = \cexpect{\bX_t}{\bX_{t-1}^{(p)} = \bxi} -f(\bxi) \label{eq:ccond}
\end{equation}
does not depend on $\bxi$ and nor, by stationarity, on $t$. But since by assumption $\bX_t$ and $\bX_{t-1}^{(p)}$ are jointly multivariate Gaussian, by a well-known result  $\cexpect{\bX_t}{\bX_{t-1}^{(p)}}$ depends linearly on $\bxi$, and from \eqref{eq:ccond} it follows that $f(\bxi)$ must be a linear function of $\bxi$.

\subsection{Invariance under transformation of variables} \label{sec:invariance}

The partial covariance $\Covop(\bX|\bY)$ transforms in a simple way under linear transformation of variables. If $T$ and $U$ are respective matrices for linear transformations on $\bX$ and $\bY$ then we have that
\begin{equation}
    \Covc{T \cdot \bX}{U \cdot \bY} \equiv T \Covc\bX\bY \trans T\,. \label{eq:pci}
\end{equation}
Using this formula, and the properties of the determinant and trace operators, we can find the respective groups of linear transformations under which  MVGC and trvMVGC are invariant. For MVGC, we find that the most general transformation that $\cgc\bY\bX\bZ$ is invariant under is given by
\begin{align}
\begin{split}
\bX &\to  T_{xx} \cdot \bX\,, \\
\bY &\to  T_{yx} \cdot \bX + T_{yy} \cdot \bY +  T_{yz} \cdot \bZ\,, \\
\bZ &\to  T_{zx} \cdot \bX + T_{zz} \cdot \bZ\,,
\end{split} \label{xmfns}
\end{align}
where the matrices $T_{xx}$, $T_{yy}$ and $T_{zz}$ on the diagonal are non-singular. All these symmetries are desirable properties for a causality measure. There ought to be invariance under redefinition of the individual variables within each of $\bX$, $\bY$ and $\bZ$, (i.e.~under the diagonal components $T_{xx}$, $T_{yy}$ and $T_{zz}$ of Eq.~\eqref{xmfns}), because MVGC is designed to measure causality between unified wholes rather than between arbitrarily defined constituent elements. The ``off-diagonal'' components $T_{yx}$, $T_{yz}$ and $T_{zx}$ are also intuitive. Adding components of $\bZ$ or $\bX$ to the predictor $\bY$ should not change the value of MVGC, because MVGC is designed to measure the ability of $\bY$ at predicting $\bX$ \textit{over and above} $\bZ$ and $\bX$. Similarly, adding components of $\bX$ onto $\bZ$ should not make a difference because the predictee $\bX$ could already be thought of as a conditional variable before transformation.

trvMVGC has an invariance under a similar group of transformations but with one significant restriction, namely that the matrix $T_{xx}$ must be \textit{conformal} (angle-preserving), that is $T_{xx}$ must satisfy $T_{xx}\trans{T_{xx}} = c I$ for some constant $c$. This difference can have practical consequences. The broader invariance of MVGC (under \emph{all} linear transformations $T_{xx}$) means that this measure, but not trvMVGC, is insensitive to certain common inaccuracies of data collection, namely those in which variables within a given set $\bX$ are contaminated by contributions from other variables (see Discussion). To put this point another way, if one wishes to infer MVGC between \emph{hidden} variables by analyzing MVGC between observed variables, these two quantities are actually the same if the relationship between hidden and observed variables is linear and can be written in the form given in Eq.~\eqref{xmfns}. One may also wish to measure the MVGC from the independent components of the predictor to the independent components of the predictee. Again, the invariance properties of MVGC mean that one does not need to explicitly find these independent components; one can simply compute MVGC between observed components. These observations indicate that MVGC takes into account correlation between variables in a principled way. We see this explicitly in Section \ref{sec:expand}.

The restriction $T_{xx}\trans{T_{xx}} = c I$ for trvMVGC further implies that an uneven \emph{rescaling} of the components of the predictee variable may change the value of $\ctgc\bY\bX\bZ$. This too has practical implications, namely that trvMVGC but not MVGC can be affected by magnitude differences in the components of $\bX$, perhaps resulting from these components reflecting underlying mechanisms that are differently amplified or differentially accessible to the measuring equipment, a common situation in many neuroscience contexts (see Discussion). This sensitivity is undesirable because causal connectivity should be based on the information content of signals (c.f.\ Section~\ref{sec:gauss}), and not on their respective magnitudes.

It is worth noting that for transfer entropy the symmetry group can be extended to include all non-singular (not necessarily linear) transformations of the predictee variable, since the entropies are invariant under such transformations.\footnote{If the predictee variable has a continuous (multivariate) distribution, we note that the Jacobian determinants in the standard change-of-variables formula for entropy calculation cancel out.} Since G-causality is essentially a linear version of transfer entropy, the former should at least be invariant under the linear subgroup of transformations.

\subsection{Expansion of Multivariate Granger Causality} \label{sec:expand}

MVGC is expandable as a sum of G-causalities over all combinations of \textit{univariate} predictor and predictee variables contained within the multivariate composites. The existence of this expansion depends on the fact that determinants are decomposable into products, and that logarithms of products are decomposable into sums of logarithms. No such decomposition exists for the logarithm of a trace, and so there is no obvious way of expanding trvMVGC into combinations of univariate components.

The expansion of MVGC is not entirely straightforward because different terms in the sum involve conditioning on the past and present of different subsets of variables. However each predictor/predictee combination appears precisely once in the sum, and each term can be explained intuitively. The general formula may be written as
\begin{equation}
    \cgc\bY\bX\bZ = \sum_{i=1}^n\sum_{\alpha=1}^m \cgc{Y_\alpha}{X_i}{\bZ\mc\bX\mc Y_1\mc\ldots\mc Y_{\alpha-1}\mc X^0_1\mc\ldots\mc X^0_{i-1}}\,, \label{eq:Fexp}
\end{equation}
where the superscript `$^0$' indicates conditioning on the present (in addition to the past) of the corresponding variables. Thus, in the term for causality from $Y_\alpha$ to $X_i$ one conditions on (i) the past of the entire multivariate conditional variable $\bZ$, (ii) the past of the entire multivariate predictee variable $\bX$, (iii) the past of all predictor variables $Y_\beta$ with $\beta < \alpha$ and (iv) the present of all predictee variables $X_j$ with $j<i$. The derivation of the expansion
\eqref{eq:Fexp} is given in Appendix \ref{expandproof}.

For the case of a multivariate predictor and a univariate predictee we have
\begin{equation}
    \gc\bY X = \gc{Y_1}X + \cgc{Y_2}X{Y_1} + \cgc{Y_3}X{Y_1\mc Y_2}+ \cdots + \cgc{Y_m}X{Y_1\mc Y_2\mc\ldots\mc Y_{m-1}}\,.
\end{equation}
This formula is consistent with the intuitive idea that the total degree to which the multivariate $\bY$ helps predict the univariate $X$ is: the degree to which $Y_1$ predicts $X$, plus the degree to which $Y_2$ helps predict $X$ over and above the information already present in $Y_1$, and so on.

For the case of a multivariate predictee and a univariate predictor we have
\begin{equation}
    \gc Y\bX = \cgc Y{X_1}\bX + \cgc Y{X_2}{\bX\mc X^0_1} + \cgc Y{X_3}{\bX\mc X^0_1\mc X^0_2}+ \cdots + \cgc Y{X_n}{\bX\mc X^0_1\mc X^0_2\mc\ldots\mc X^0_{n-1}}\,.
\end{equation}
This formula supports the intuition that the total degree to which the univariate $Y$ helps predict the multivariate $\bX$ is: the degree to which the past of $Y$ helps predict the current value of $X_1$ over and above the degree to which the past of the whole of $\bX$ predicts the current value of $X_1$, plus the degree to which the past of $Y$ helps predict the current value of $X_2$ over and above the degree to which the past of the whole of $\bX$ and the current value of $X_1$ predicts the current value of $X_2$, and so on.

We remark on two implications of the
expansion of MVGC. First, Ladroue and colleagues suggested that use
of generalized residual variance for causal inference on
high-dimensional data might suffer from problems of numerical
stability. However, the expansion of MVGC into low-dimensional,
univariate G-causalities suggests that there should be no
problem (see Section \ref{sec:stability} for numerical evidence of this).
Second, the expansion \eqref{eq:Fexp} indicates that MVGC
controls for, to some extent, the influence of unmeasured
latent/exogenous variables (see also Section \ref{sec:partial}). By conditioning on the present of
certain appropriate predictee variables for each term of the
expansion, only the effects of each predictor on independent
components of the predictees enter the equation. This property stems
from the fact that the determinant of the residual covariance matrix
reflects not just residual variances, but also the extent to which
these residual variances are independent of each other. This is
another advantage of the MVGC measure over trvMVGC, which does not
depend on residual correlations.

\subsubsection{Stability of Multivariate Granger Causality} \label{sec:stability}

\begin{figure*}
\includegraphics[width=0.48\textwidth]{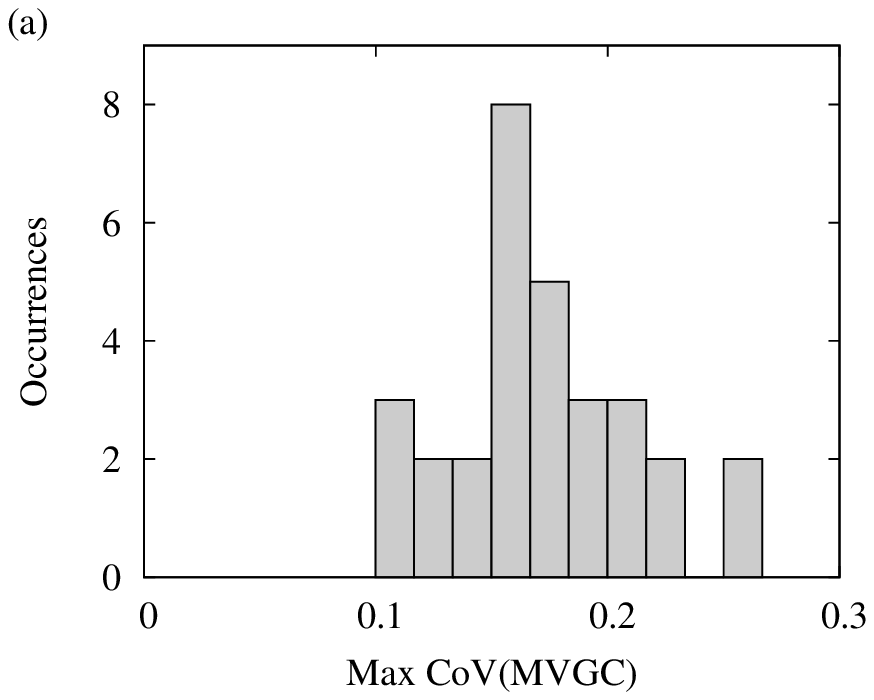} \ \ \ \
\includegraphics[width=0.48\textwidth]{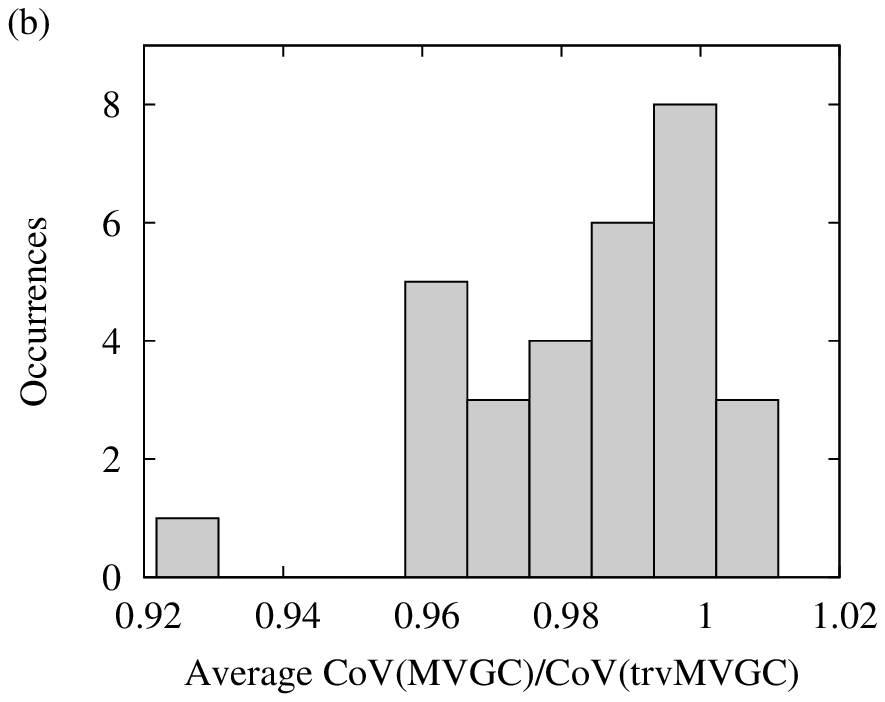}
\caption{Stability of MVGC. (a) Histogram of the maximum CoV of MVGC, observed over 10 trials of 3000 time-steps, for each of 30 different systems, as described in Section \ref{sec:stability}. (b) Histogram of the average ratio between the CoV of MVGC and the  CoV of trvMVGC, for each of the 30 systems. MVGC is numerically stable (a) and is not appreciably different from trvMVGC in terms of stability properties (b). \label{fig:stability1}}

\end{figure*}

We tested numerically our claim (Section \ref{sec:expand}) that MVGC should not be less stable than trvMVGC. We studied MVAR$(1)$ processes whose dynamics are given by
\begin{equation} \label{stableprocess}
\bX_t=A\cdot \bX_{t-1} +\beps_t\,,
\end{equation}
where $\bX$ contains 8 variables, the sum of each row of $A$ (i.e.~total afferent to each element) is 0.5, all components in a given row of $A$ are equal and positive, and each component of $\beps_t$ is an independent Gaussian random variable of mean 0 and variance 1. We generated 30 random ``connectivity'' matrices (or systems) $A_i$, $(i=1,\ldots,30)$, each with an average of 2 non-zero components per row. For each $A_i$ we obtained 10 sets of 3000 (post equilibrium) data points via Eq.~\eqref{stableprocess}. For each set, we computed the MVGC across each bipartition of the system corresponding to $A_i$.  We then calculated, for each bipartition, the standard deviation of the MVGC across the 10 data sets and (excluding bipartitions with standard deviation less than 0.01) the corresponding coefficient of variation (CoV, standard deviation divided by mean).  This procedure allowed us to obtain, for each $A_i$, a maximum CoV. Figure \ref{fig:stability1}(a) shows that the maximum CoV is generally very small and never large, confirming the stability of MVGC.

To compare the stability of MVGC with that of trvMVGC, for each $A_i$ and for each bipartition we divided the CoV for MVGC by the CoV for trvMVGC. Figure \ref{fig:stability1}(b) shows the distribution of the average of this ratio across all bipartitions.  The clustering of this distribution at $\approx$1, with no outliers, confirms that MVGC and trvMVGC have similar stability properties, at least in the systems we have simulated.

To generalize these results we next used a genetic algorithm (GA) \cite{Seth05,SpornsEtal00} to see if we could find a network for which MVGC becomes unstable. The GA was initialized using a population composed of the 30 random systems $A_i$ described above. We ran the GA for 130 generations. In each generation, we computed the fitness of each system as the maximum CoV of MVGC. Systems were selected to proceed to subsequent generations using stochastic rank-based selection. Mutations enabled the adding of new non-zero components to $A_i$, the removal of existing non-zero components, or the swapping of components, followed by renormalization of each row to sum to 0.5 again; two mutations were applied per system.  After 130 generations (sufficient for fitness to asymptote) the average fitness (i.e.~maximum CoV) in the population was $\approx$0.25, and the maximum was 0.39, which is still a low value.  For the $A_i$ that gave this highest value, we compared the CoV obtained using MVGC with that obtained using trvMVGC following the procedure described above.  The average ratio (across all bipartitions) was $\approx$1.00, (maximum value 1.12), indicating that MVGC and trvMVGC had similar stability properties even for systems optimized to be unstable with respect to MVGC.  Further, we examined some $A_i$ for which the sums of the rows differed (i.e.~having heterogeneous afferent connectivity); these systems had similar stability properties to those described above. Finally, stability properties were unaffected when computations were based on 1000 (rather than 3000) data points.

Taken together, these simulation results confirm that MVGC is numerically stable, and is not appreciably different from trvMVGC in terms of stability properties.

\subsection{Spectral decomposition} \label{sec:sdecomp}

In this section we review the spectral decomposition of G-causality \cite{Geweke82,DingEtal06}. For simplicity we limit ourselves to the unconditional case, although the procedure may be readily extended to the conditional case (as described in \eg\ Refs.~\cite{Geweke84,DingEtal06,Chen06}).  We assume multivariate predictor and predictee variables, and show that MVGC but not trvMVGC has a satisfactory spectral decomposition.

Consider the stationary MVAR
\begin{equation}
    \bX_t = A \cdot \bX^{(p)}_{t-1} + \beps_t = \sum_{k=1}^p  A_k \cdot \bX_{t-k} + \beps_t\,. \label{eq:mvar}
\end{equation}
We may write this as
\begin{equation}
     A(L) \cdot \bX_t= \beps_t\,, \label{eq:alag}
\end{equation}
where $L$ denotes the (single time step) lag operator, and
\begin{equation}
    A(L) \equiv I - \sum_{k=1}^p A_k L^k\,.
\end{equation}
Eq.~\eqref{eq:alag} may be solved as
\begin{equation}
    \bX_t = H(L) \cdot \beps_t\,, \label{eq:hlag}
\end{equation}
where $H(L) \equiv A(L)^{-1}$. Transforming into the frequency domain via the discrete-time Fourier transform $\bX(\lambda) = \sum_{t = -\infty}^\infty \bX_t \;e^{-i\lambda t}$ yields $ A(\lambda) \cdot \bX(\lambda) = \beps(\lambda)$ (replace $L$ by $e^{-i\lambda}$), so that
\begin{equation}
    \bX(\lambda) = H(\lambda) \cdot \beps(\lambda)\,,
\end{equation}
where $H(\lambda) \equiv A(\lambda)^{-1}$ is the \emph{transfer matrix}. The (power) \emph{spectral density} of $\bX$ is then given by
\begin{equation}
    S(\lambda) = H(\lambda) \Covs{\beps} \ctran H(\lambda)\,. \label{eq:sh}
\end{equation}
From a standard result \cite{Rozanov67}, since $H(L)$ is a square matrix lag operator with the identity matrix as leading term, we have
\begin{equation}
    \frac 1{2\pi} \int_{-\pi}^\pi \ln\dett{H(\lambda) \ctran H(\lambda)} \dlam = 0\,, \label{eq:hint}
\end{equation}
provided that all roots of the characteristic polynomial $\dett{A(L)}$ lie outside the unit circle, which is a necessary condition for the existence of the stationary process \eqref{eq:mvar}. From \eqref{eq:sh} we may then derive the relation \cite{Doob53}
\begin{equation}
    \frac 1{2\pi} \int_{-\pi}^\pi \ln\dett{S(\lambda)} \dlam = \ln\dett{\Covs\beps}\,. \label{eq:sint}
\end{equation}

Consider now the stationary MVAR
\begin{equation}
    \bX_t \mc\bY_t  = A \cdot \bracr{\bX^{(p)}_{t-1} \mc \bY^{(q)}_{t-1}} + \beps_{x,t} \mc \beps_{y,t} \label{eq:xymvar}
\end{equation}
with coefficients matrix
\begin{equation}
    A \equiv
    \begin{pmatrix}
        A_{xx} & A_{xy} \\
        A_{yx} & A_{yy}
    \end{pmatrix}\,
\end{equation}
and residuals covariance matrix
\begin{equation}
    \Covs{\beps_x \mc \beps_y} \equiv
    \begin{pmatrix}
        \Sigma_{xx} & \Sigma_{xy} \\
        \Sigma_{yx} & \Sigma_{yy}
    \end{pmatrix}\,.
\end{equation}
Let us split the corresponding transfer matrix $H(\lambda)$ as
\begin{equation}
    H(\lambda) \equiv A(\lambda)^{-1} =
    \begin{pmatrix}
        H_{xx}(\lambda) & H_{xy}(\lambda) \\
        H_{yx}(\lambda) & H_{yy}(\lambda)
    \end{pmatrix}\,
\end{equation}
and the spectral density as
\begin{equation}
    S(\lambda) =
    \begin{pmatrix}
        S_{xx}(\lambda) & S_{xy}(\lambda) \\
        S_{yx}(\lambda) & S_{yy}(\lambda)
    \end{pmatrix}\,.
\end{equation}
Then $S_{xx}(\lambda)$ is just the spectral density of $\bX$, which from \eqref{eq:sh} is given by
\begin{equation}
    S_{xx}(\lambda)  =  H_{xx}(\lambda) \Sigma_{xx} \ctranx H{xx}(\lambda) \\
      + 2\,\re\!\bracc{H_{xx}(\lambda) \Sigma_{xy} \ctranx H{xy}(\lambda)} \\
      + H_{xy}(\lambda) \Sigma_{yy} \ctranx H{xy}(\lambda)\,. \label{eq:sxx}
\end{equation}
The idea is that we wish to decompose this expression into a part reflecting the effect of $\bX$ itself and a part reflecting the causal influence of $\bY$. The problem is that, due to the presence of the ``cross'' term, $S_{xx}(\lambda)$ does not split cleanly into an $\bX$ and a $\bY$ part. Geweke \cite{Geweke82} addresses this issue by introducing the transformation
\begin{equation}
    \bX\mc\bY \to U \cdot \bracr{\bX\mc\bY}\,,
\end{equation}
where
\begin{equation}
    U \equiv \begin{pmatrix} I & 0 \\ -\Sigma_{yx} \Sigma_{xx}^{-1} & I \end{pmatrix}\,. \label{eq:cxform}
\end{equation}
Note that this transformation leaves the G-causality $\gc\bY\bX$ invariant (c.f.~Section \ref{sec:invariance}) and, for the transformed regression, we have $\Sigma_{xy} \equiv 0$; that is, the residuals $\beps_x, \beps_y$ are uncorrelated. Thus, assuming the transformation \eqref{eq:cxform} has been pre-applied, Eq.~\eqref{eq:sxx} becomes
\begin{equation}
    S_{xx}(\lambda) = H_{xx}(\lambda) \Sigma_{xx} \ctranx H{xx}(\lambda) + H_{xy}(\lambda) \Sigma_{yy} \ctranx H{xy}(\lambda)\,, \label{eq:sxxt}
\end{equation}
whereby the spectral density of $\bX$ splits into an ``intrinsic'' part and a ``causal'' part. The spectral G-causality of $\bY \to \bX$ at frequency $\lambda$ is now defined to be
\begin{equation}
    \sgc\bY\bX(\lambda) \equiv \lnt{\frac{\dett{S_{xx}(\lambda)}}{\dett{H_{xx}(\lambda) \Sigma_{xx} \ctranx H{xx}(\lambda)}}} \label{eq:sgc}
\end{equation}
or, in terms of the \emph{untransformed} variables,
\begin{equation}
    \sgc\bY\bX(\lambda) \equiv \lnt{\frac{\dett{S_{xx}(\lambda)}}{\dett{S_{xx}(\lambda) - H_{xy}(\lambda) \Sigma_{y|x} \ctranx H{xy}(\lambda)}}}\,, \label{eq:sgcut}
\end{equation}
with $S_{xx}(\lambda)$ as in \eqref{eq:sxx} and $\Sigma_{y|x} \equiv \Sigma_{yy} - \Sigma_{yx} \Sigma_{xx}^{-1} \Sigma_{xy}$.

Geweke (Ref.~\cite{Geweke82}, Theorem 2) then establishes the fundamental motivating relationship between frequency and time domain G-causality:
\begin{equation}
    \frac 1{2\pi} \int_{-\pi}^\pi \sgc\bY\bX(\lambda) \dlam = \gc\bY\bX\,, \label{eq:sgcint}
\end{equation}
provided that all roots of $\dett{A_{yy}(L)}$ lie outside the unit circle.\footnote{A subtlety to note is that even if the MVAR \eqref{eq:xymvar} has a finite number of lags $p,q < \infty$, the exact \emph{restricted} regression of $\bX$ on its own past will generally require an infinite number of lags \cite{Geweke82}. Thus in theory, for exact equality in \eqref{eq:sgcint}, an infinite number of lags is required to calculate the term $\Covc{\bX}{\blX}$ which appears in $\gc\bY\bX$ (using a finite number of lags will generally result in an overestimate of $\gc\bY\bX$, since residual errors will be larger than for the exact regression). As applied to empirical data, it is in any case good practice to choose ``sufficient'' lags for all regressions so as to model the data adequately without overfitting \cite{Akaike74,Schwartz78}.\label{fn:gcint}} The proof of this relation relies crucially on the result \eqref{eq:hint} which, we note, involves the \emph{determinant} of the transfer matrix. Thus if the trace, rather than the determinant, were to be used in the definition \eqref{eq:sgc} for $\sgc\bY\bX(\lambda)$ then we could not expect to obtain a relation corresponding to \eqref{eq:sgcint}, since (i) the trace of the spectral density in Eq.~\eqref{eq:sh} does not factorize, (ii) there is no trace analogue to Eq.~\eqref{eq:hint}, and thus (iii) no analogue to Eq.~\eqref{eq:sint}. This would seem to preclude a satisfactory spectral decomposition for the trace version of G-causality. Similar remarks apply to conditional G-causality in the spectral domain.

\begin{figure*}
\includegraphics[width=0.48\textwidth]{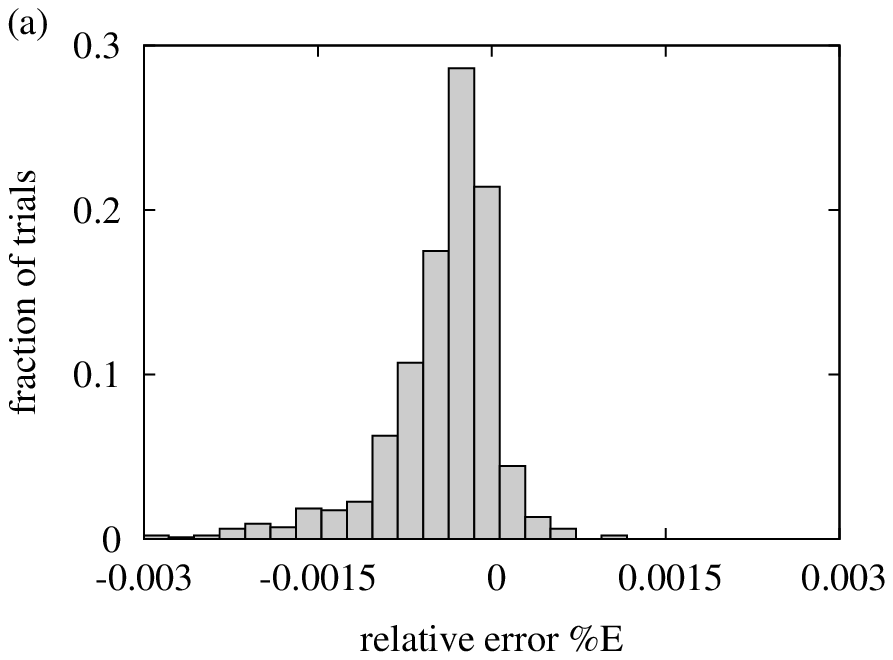} \ \ \ \
\includegraphics[width=0.48\textwidth]{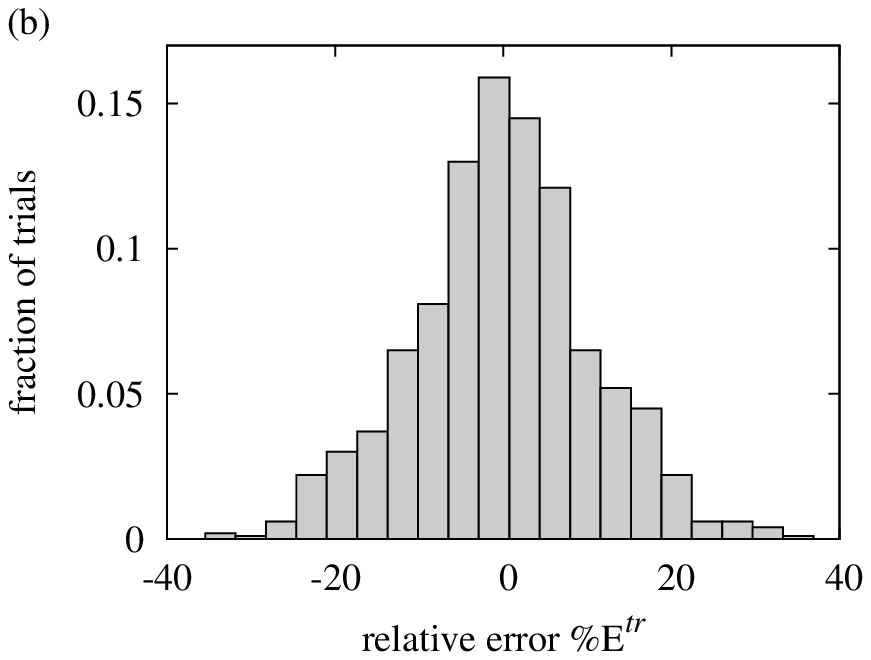}
\caption{Distribution of relative errors of integrated spectral multivariate G-causality with respect to the time domain for (a) MVGC (b) trvMVGC, for a random sample of MVAR$(1)$ processes.} \label{fig:speccomp_hist}
\end{figure*}

In Ref.~\cite{ladroue:2009}, however, it is conjectured that a trace analogue of Eq.~\eqref{eq:sgcint} does indeed hold. To test this conjecture we performed the following experiment: we simulated $1000$ MVAR$(1)$ processes of the form
\begin{equation}
    \bX_t \mc Y_t = A \cdot \bracr{\bX_{t-1} \mc Y_{t-1}} + \beps_{x,t} \mc \eps_{y,t}\,, \label{eq:mvar1}
\end{equation}
where $\bX$ has dimension $2$ and $Y$ dimension $1$. Residuals $\beps_{x,t}, \eps_{y,t}$ were completely uncorrelated, with unit variance (\ie\ $\Covs{\beps_{x,t} \mc \eps_{y,t}}$ was the $3 \times 3$ identity matrix) so that, in particular, the Geweke transformation \eqref{eq:cxform} was unnecessary. For each trial the $3 \times 3$ coefficients matrix $A$ was chosen at random with elements uniform on $[-\shalf,\shalf]$, and the process \eqref{eq:mvar1} simulated for $10^6$ stationary time steps (the occasional unstable process was rejected). Time domain causalities $\gc Y\bX, \ \tgc Y\bX$ and frequency domain causalities $\sgc Y\bX(\lambda), \ \tsgc Y\bX(\lambda)$ were calculated in sample using $p = 10$ lags. (As noted previously,$^{10}$ equality in \eqref{eq:sgcint} is only assured in the limit of infinite lags; $10$ lags was found empirically to achieve good accuracy  without overfitting the data.) Relative errors of integrated spectral MVGC with respect to time-domain MVGC, expressed as a percentage, were defined as
\begin{align}
	E_\%\phantom{^{tr}} &\equiv 100 \times \frac{\frac 1{2\pi} \int_{-\pi}^\pi \sgc Y\bX(\lambda) \dlam-\gc Y\bX}{\gc Y\bX}\,, \nonumber \\
	E_\%^{tr} &\equiv 100 \times \frac{\frac 1{2\pi} \int_{-\pi}^\pi \tsgc Y\bX(\lambda) \dlam-\tgc Y\bX}{\tgc Y\bX}\,,
\end{align}
for MVGC and trvMVGC respectively. (The integrals were computed by standard numerical quadrature.) Results, displayed in Table~\ref{tab:speccomp}, confirm to good accuracy the theoretical prediction of Eq.~\eqref{eq:sgcint} for MVGC (the small negative bias on $E_\%$ is due to the finite number of lags), while for trvMVGC relative errors are several orders of magnitude larger and furthermore were not decreased by choosing longer stationary sequences and/or more lags.
\begin{table}
\begin{center}
{
\renewcommand{\arraystretch}{1.5}
\begin{tabular}{|c|c|c|c|}
 \hline
error & mean & std.~dev. & abs.~mean  \\
\hline
$E_\%\phantom{^{tr}}$ & $-0.0004$ & $\phantom 1 0.0005$ & $0.0005$ \\
\hline
$E_\%^{tr}$ & $-0.0488$ & $10.5995$ & 8.1799 \\
\hline
\end{tabular}
}
\caption{Comparison of relative errors of integrated spectral MVGC and trvMVGC with respect to time domain MVGC and trvMVGC, for a random sample of MVAR$(1)$ processes. Top row shows MVGC, bottom row shows trvMVGC. See text for details. Figures in the ``abs.~mean'' column are the means of the absolute values $\abs{E_\%}$ and $\abs{E_\%^{tr}}$.} \label{tab:speccomp}
\end{center}
\end{table}
The full distribution of relative errors is also displayed as a histogram in Fig.~\ref{fig:speccomp_hist}.

\begin{figure*}
\includegraphics[width=0.48\textwidth]{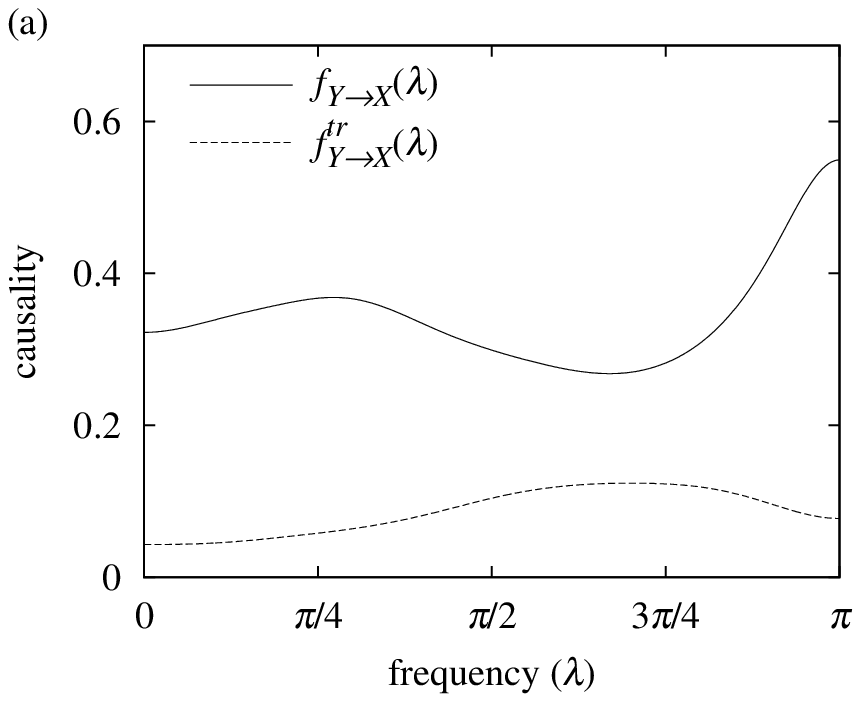} \ \ \ \
\includegraphics[width=0.48\textwidth]{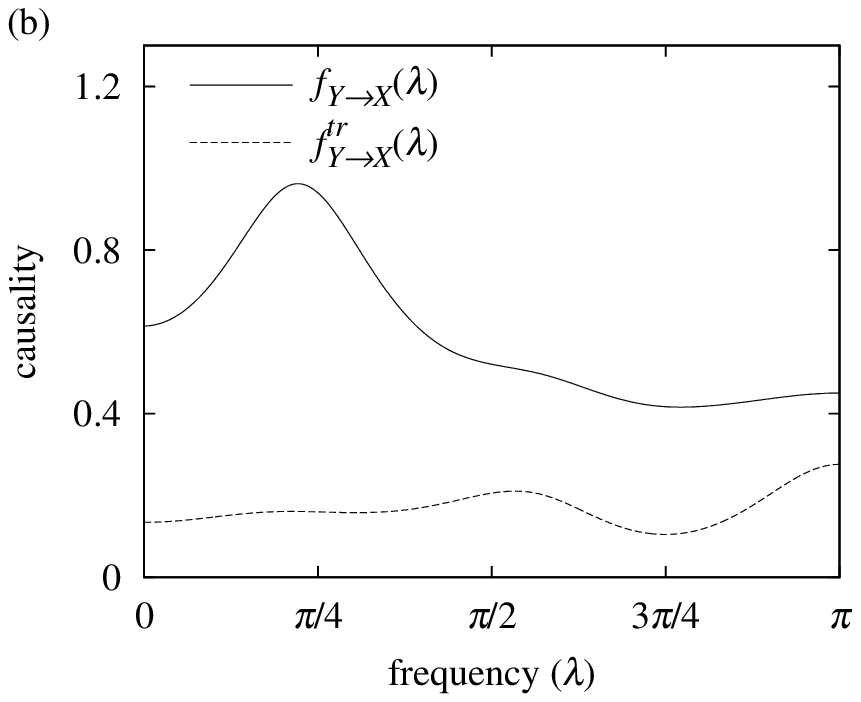}
\caption{Comparison of MVGC and trvMVGC in the frequency domain: spectral MVGC and trvMVGC plotted against frequency for (a) a typical MVAR$(3)$ process with $\dim(\bX) = 2, \ \dim(Y) = 1$ and (b) a typical MVAR$(5)$ process with $\dim(\bX) = 3, \ \dim(\bY) = 2$.} \label{fig:speccomp_single}
\end{figure*}

We also repeated the experiment with higher order MVAR$(p)$ processes, higher dimensional predictee and predictor variables and correlated residuals $\beps_x$. In all cases, results confirmed the accuracy of \eqref{eq:sgcint} for MVGC and yielded large relative errors for trvMVGC. We remark that \emph{qualitative} differences (\ie\ aside from differences of scale) between spectral MVGC and trvMVGC could be substantial (Fig.~\ref{fig:speccomp_single}). These differences, furthermore, appeared in general to be exaggerated by the presence of residual correlations; this is consonant with the sensitivity of MVGC as contrasted with the lack of sensitivity of trvMVGC to residual correlations (see Sections \ref{sec:expand} and \ref{sec:partial}).

It is straightforward to show that $\sgc\bY\bX(\lambda)$ is invariant under the same group of linear transformations \eqref{xmfns} as $\gc\bY\bX$; again, $\tsgc\bY\bX(\lambda)$ will in general be invariant only under the restricted group with $T_{xx}$ conformal; this extends to the conditional case.

\section{Multivariate partial Granger causality} \label{sec:partial}

Recently, a \textit{partial G-causality} measure has been introduced \cite{guoetal:2008} which exploits a parallel with the concept of \textit{partial coherence} \cite{baccala:2001} in order to control for latent/exogenous influences on standard G- causality. Partial G-causality modifies the standard G-causality measure by including terms based on residual correlations between the predictee variable and the conditional variables. Consider, in addition to the regressions \eqref{eq:reg}, the following regressions of the \textit{conditioning} variable $\bZ_t$:
\begin{equation}
\begin{split}
    \bZ_t & = B  \cdot \bracr{\bX^{(p)}_{t-1}\mc\bZ^{(r)}_{t-1}}  + \beeta_t\,, \\
    \bZ_t & = B' \cdot \bracr{\bX^{(p)}_{t-1}\mc\bY^{(q)}_{t-1}\mc\bZ^{(r)}_{t-1}} + \beeta'_t\,.
\end{split} \label{eq:regz}
\end{equation}
Here the roles of the predictee and conditioning variables are reversed. Then for univariate predictor and predictee the partial G-causality of $Y$ on $X$ given $\bZ$ is defined by conditioning the respective residual covariances for the regressions of $X$ on the corresponding residuals for the regressions of $\bZ$:
\begin{equation}
    \cpgc Y X\bZ \equiv \lnt{\frac{\Covc{\beps_t}{\beeta_t}}{\Covc{\beps'_t}{\beeta'_t}}}\,. \label{eq:pgcue}
\end{equation}
This extends naturally to the fully multivariate case (c.f.~Eq.~\eqref{eq:gc}), and we define partial MVGC (pMVGC) as
\begin{align}
	\cpgc\bY\bX\bZ
	&\equiv \lnt{\frac{\dett{\Covc{\beps_t}{\beeta_t}}}{\dett{\Covc{\beps'_t}{\beeta'_t}}}} \label{eq:pgcd} \\
	&= \lnt{\frac{\dett{\Covc{\bX}{\blX\mc\blZ\mc\bZ}}}{\dett{\Covc{\bX}{\blX\mc\blY\mc\blZ\mc\bZ}}}} \label{eq:pgc}
\end{align}
where the RHS \eqref{eq:pgc} follows from the identity \eqref{eq:rcepet} derived in Appendix \ref{apx:pgc}, (with $\bW \equiv \blX\mc\blZ$ and $\bW \equiv \blX\mc\blY\mc\blZ$ for the numerator and denominator terms respectively). Comparing with \eqref{eq:gc} we see thus that pMVGC differs from MVGC in the inclusion of the \textit{present} of the conditioning variable $\bZ$ in the respective regressions. Seen in this form, it is clear that, as is the case for MVGC, pMVGC is always non-negative.\footnote{In \cite{guoetal:2008} it is stated that partial G-causality may in some circumstances be \textit{negative}; the justification for this is unclear.} One could alternatively express pMVGC as (non-partial) MVGC conditioned on a ``forward lagged'' version of $\bZ$: defining $\btZ_t \equiv \bZ_{t+1}$ we have $\bZ_t\mc\bZ^{(r)}_{t-1} \equiv \btZ^{(r+1)}_{t-1}$, or $\bltZ = \bZ\mc\blZ$ (note the additional lag on $\bltZ$), so that, from Eq.~\eqref{eq:pgc},
\begin{equation}
    \cpgc\bY\bX\bZ = \cgc\bY\bX{\tilde\bZ} \label{eq:pgcz1}\,.
\end{equation}

As noted in Section \ref{sec:expand}, (non-partial) MVGC to some extent already controls for the influence of latent/exogenous variables because the generalized variance is sensitive to residual correlations.  However,  pMVGC takes into account even more correlations with the explicit aim of controlling for latent/exogenous influences. pMVGC may therefore be preferable when such influences are expected to be (a) strong and (b) relatively uniform in their influence on the measured system.  Indeed, pMVGC (and the original measure of partial G-causality) can only be effective in compensating for latent/exogenous variables that affect \textit{all} modeled variables (\ie\ predictee, predictor and conditioning) to a roughly equal degree \cite{guoetal:2008}.

It is interesting to note that pMVGC may be expressed in terms of non-partial MVGCs as
\begin{equation}
    \cpgc\bY\bX\bZ = \gc\bY{\bZ\mc\bX}-\cgc\bY\bZ\bX\,. \label{eq:pgc1}
\end{equation}
by straightforward application of Eq.~\eqref{eq:cxyid}. As expected, \eqref{eq:pgc1} includes a term with a mandatory multivariate predictee, since it is only in this case that residual correlation can make a difference. It is interesting that $\bZ$ appears as a \textit{predictee} variable; this might be understood as  pMVGC using the conditioning variable $\bZ$ as a ``proxy'' by which to assess the influence of latent or exogenous variables.

A ``trace'' version of pMVGC may be defined analogously to \eqref{eq:pgcd}. Again by Eq.~\eqref{eq:rcepet} of Appendix \ref{apx:pgc}, the identity corresponding to \eqref{eq:pgc} will hold, as will the trace analogue of \eqref{eq:pgcz1}. However, the analogue of \eqref{eq:pgc1} will \emph{not} hold in general, since the traces of the partial covariance matrices will in general not factorize appropriately.\footnote{In \cite{ladroue:2009}, under the section headed ``Partial Complex Granger causality'', the quantity developed appears to be (the trace version of) what is conventionally referred to as \emph{conditional} G-causality, rather than partial G-causality as introduced in \cite{guoetal:2008} and referenced in this section.}

From \eqref{eq:pgcz1} it is straightforward to derive a spectral decomposition $\pscgc\bY\bX\bZ(\lambda)$ for pMVGC, which will integrate correctly to the time-domain pMVGC $\cpgc\bY\bX\bZ$. Again, a spectral decomposition for the corresponding trace version is likely to be problematic, insofar as it will fail in general to integrate correctly to the time-domain value (c.f.~Section \ref{sec:sdecomp}).

\section{Causal density}

A straightforward application of MVGC is to measures of
\textit{causal density}, the overall level of causal interactivity
sustained by a multivariate system $\bX$. A previous measure of
causal density \cite{Seth05} has been defined as the average of all
pairwise (and hence univariate) G-causalities between system
elements, conditioned on the remaining system
elements:\footnote{This is the ``weighted'' version of causal
density. An unweighted and [0,1] bounded alternative can be defined
as the fraction of all pairwise conditional causalities that are
statistically significant at a given significance level.}
\begin{equation}
    \cd\bX \equiv \frac 1{n(n-1)} \sum_{i \ne j} \cgc{X_i}{X_j}{\bX_{[ij]}} \label{cd1}
\end{equation}
where $\bX_{[ij]}$ denotes the subsystem of $\bX$ with variables $X_i$ and $X_j$ omitted, and $n$ is the total number of variables.  Causal density provides a useful measure of the dynamical ``complexity'' of a system inasmuch as elements that are completely independent will have zero causal density, as will elements that are completely integrated in their dynamics. Exemplifying standard intuitions about complexity \cite{sporns:2007}, high causal density will only be achieved when elements behave somewhat differently from each other, in order to contribute novel potential predictive information, and at the same time are globally integrated, so that the potential predictive information is in fact useful \cite{SethEtal06,shanahan:2008}.

Using MVGC, various extensions to \eqref{cd1} can be suggested, based on the various possible interactions between multivariate predictors, predictees and conditional variables. These extensions may provide a more principled measure of complexity by analyzing a target system at multiple scales. First we define the causal density from size $k$ to size $r$, $\mathrm{cd}_{k\to r}(\boldsymbol{X})$, as the average MVGC from a subset of size $k$ to a subset of size $r$, conditioned on the rest of the system:
\begin{equation}
	\cdx{k\to r}\bX = \frac 1{n_{k,r}}\sum_{i=1}^{n_{k,r}} \cgc{\bV^k_i}{\bU^r_i}{\bW^{n-k-r}_i}\,,
\end{equation}
where $\bX = \bV^k_i \cup \bU^r_i \cup \bW^{n-k-r}_i$ denotes the $i^\mathrm{th}$ of the $n_{k,r} \equiv \binom n k \binom{n-k}r$ distinct tripartitions of $\bX$ into disjoint sub-systems of respective sizes $k$, $r$ and $(n-k-r)$. Then using this, one could define the \textit{bipartition causal density} (bcd) as the average of $\mathrm{cd}_{k\to (n-k)} (\bX)$ over predictor size $k$,
\begin{equation}
	\bcd\bX = \frac{1}{n-1} \sum_{k=1}^{n-1} \cdx{k \to (n-k)}\bX \,.
\end{equation}
Interestingly, this quantity is closely related to the popular Tononi-Sporns-Edelman ``neural complexity'' measure \cite{TononiEtal94} which averages (contemporaneous) \textit{mutual information} across bipartitions; (we are currently exploring this in work in preparation). It could also be interesting to compare causal density at different scales of predictor plus predictee size; thus we define
\begin{equation}
	\cdx s\bX \equiv \frac 1{s-1} \sum_{k=1}^{s-1} \cdx{k \to (s-k)}\bX\,.
\end{equation}
Then the original causal density measure of Eq.~\eqref{cd1} is just $\mathrm{cd}_2$ and bcd is $\mathrm{cd}_n$. The average of this over all scales can be used to define a complete \textit{tripartition causal density} (tcd):
\begin{equation}
	\tcd\bX \equiv \frac 1{n-1} \sum_{s=2}^n \cdx s\bX\,.
\end{equation}
A comparison of the properties of all versions of causal density, as well as related complexity measures, is in progress. We remark that it is straightforward to define spectral versions of these causal density measures.
\\

\section{Autonomy in complex systems}

G-causality has recently been adapted to provide an operational measure of ``autonomy'' in complex systems \cite{seth:alife:2009}. A variable $\bX$ can be said to be ``G-autonomous'' with respect to a (multivariate) set of external variables $\bZ$ if its own past states help predict its future states over and above predictions based on $\bZ$.  This definition rests on the intuition of autonomy as ``self determination'' or ``self causation''. We can formalize this notion along the lines of MVGC as follows.  Consider the regressions
\begin{equation}
\begin{split}
    \bX_t & = A  \cdot \bZ^{(r)}_{t-1}  + \beps_t\,, \\
    \bX_t & = A' \cdot \bracr{\bX^{(p)}_{t-1}\mc\bZ^{(r)}_{t-1}} + \beps'_t\,,
\end{split} \label{eq:maut}
\end{equation}
which differ from Eqs.~\eqref{eq:reg} primarily because the predictee variable $\bX$ is \emph{not} regressed on itself in one of the equations. The G-autonomy of $\bX$ is then given by
\begin{equation}
    \mathcal{A}_{\bX|\bZ} = \lnt{\frac{\dett{\Covs{\beps_t}}}{\dett{\Covs{\beps'_t}}}}\,. \label{eq:aut}
\end{equation}
The extension of G-autonomy to the multivariate case is important because it accommodates situations in which groups of elements may be jointly autonomous (self-determining, self-causing), even though the activity of individual elements within the group may be adequately predicted by combinations of activities of other elements in the group.  Univariate formulations of G-autonomy \cite{seth:alife:2009} would fail in these cases. Consider as a trivial example an element $X_1$ which is G-autonomous with respect to a background $\bZ$.  If $X_1$ is now duplicated by the element $X_2$ it will no longer appear as G-autonomous within the multivariate system $X_1{\mc}X_2{\mc}\bZ$.  However, the multivariate variable $X_1{\mc}X_2$ will be (jointly) G-autonomous with respect to $\bZ$.

As discussed in \cite{seth:alife:2009} G-autonomy also provides
the basis for a notion of ``G-emergence'' as applied to the relation
between \emph{macroscopic} variables ``emerging'' from the activity of
\emph{microscopic} constituents.  G-emergence operationalizes the
intuition that a macro-level variable is emergent to the extent that
it is simultaneously \emph{autonomous from} and \emph{dependent
upon} its micro-level constituents \cite{seth:alife:2009,Bedau97}.
Extension of G-emergence to the multivariate case using MVGC is
straightforward, allowing consideration of multivariate micro- and
macro-variables.

\section{Macroscopic variables and causal independence}

Given the ability to assess multivariate causal interactions, a
second challenge arises: the identification of relevant groupings of
variables into multivariate ensembles. One approach to this challenge
adopts the perspective of statistical mechanics on the emergence of novel macroscopic variables,
given a microscopic description of a system
\cite{ShaliziMoore06,shalizi:2006}. Here, we suggest that MVGC may
furnish a useful method for macro-variable identification in this
context.  Let us assume that $\bZ_t$ represents a set of
microscopic variables defining a complex (possibly stochastic) dynamical system, and
$\bX_t \equiv f(\bZ_t)$ a set of macroscopic variables functionally
(possibly deterministically) dependent on the microscopic variables.
There is then a sense in which $\bX$ represents a ``parsimonious''
high-level description of the system, to the extent that it predicts
its own dynamical evolution without recourse to the low level of
description of the system represented by $\bZ$; that is, to the
extent that $\bX$ exhibits strong \emph{causal independence} with
respect to $\bZ$. In this view, $\gc\bZ\bX$ furnishes a natural
measure of the \emph{lack} of this causal independence, which might
then be used to \emph{identify} parsimonious macroscopic variables
by minimizing $\gc\bZ{f(\bZ)}$ over candidate functions $f(\cdot)$.
The multivariate formulation MVGC would appear to be
significant in this context for reasons similar to the G-autonomy
case. Specifically, it may be that a set of macroscopic variables
$\bX$ may \emph{jointly} have high causal independence with respect
to the microscopic variables $\bZ$, while the component variables
$X_i$ may individually have lower causal independence.

The notions of G-autonomy, G-emergence, and causal independence are distinct but related. In short, G-autonomy measures ``self-causation'', causal independence measures the \emph{absence} of useful predictive information between microscopic and macroscopic descriptions of a system, and G-emergence measures a combination of macro-level autonomy and micro-to-macro causal \emph{dependence}.  It is possible, and is left as an objective of future work, that all three measures could be applied usefully to systems that avail multiple levels of descriptions, (i) to identify relevant groupings of observables at each level, (ii) to decompose causal interactions within each level, and finally (iii) to quantitatively characterize inter-level relationships.

\section{Discussion}

We have described and motivated a measure of multivariate causal
interaction that is a natural extension of the standard G-causality measure.  The measure, originally introduced by Geweke
\cite{Geweke82} but almost totally overlooked since, uses the
generalized variance (the determinant of the residual covariance
matrix) and we have termed it \emph{multivariate G-causality}
(MVGC). It contrasts with another recent proposal
\cite{ladroue:2009} for addressing the same problem which uses
instead the total variance (the trace of the residual covariance
matrix).  In this paper, we have presented several theoretical
justifications, augmented by numerical modeling, for preferring MVGC over the trace version, which we
summarize below. We have also extended MVGC to address novel
challenges in the analysis of complex dynamical systems, including
quantitative characterization of ``causal density'', ``autonomy'', and
the identification of novel macroscopic variables via causal
independence.

\subsection{Importance of multivariate causal analysis}

In many analyses of complex systems, particularly in neuroscience
and biology, there may be no simple or principled relationship
between observed variables and explanatorily relevant
collections, or ensembles, of these variables.  In the Introduction
we already remarked on fMRI, where
explanatorily relevant ROIs are each composed of multiple
observables (voxels) which are arbitrarily demarcated with respect to underlying neural mechanisms. Other non-invasive neuroimaging methods share similar varieties of arbitrariness:  both electroencephalography (EEG) and magnetoencephalography (MEG) provide signals which are complex convolutions of underlying neural sources.  In these and similar cases, multivariate causal analysis, and MVGC in particular, can be used to aggregate univariate observables into meaningful multivariate (ensemble) variables.  It bears emphasizing that MVGC is fundamentally different from conditional G-causality \cite{Seth07a}, which assesses the causal connectivity between two univariate variables, conditioned on a set of other variables.

Even when it is possible to measure directly the activity of variables of interest, it is still important to consider multivariate interactions.  Continuing with the neuroscience example, it may be that multiple ROIs act \emph{jointly} to influence other ROIs, or cognitive and/or behavioral outputs.  In single cell recordings this point is even more pressing: since Hebb \cite{Hebb49} it has been increasingly appreciated that neurons act as ensembles, rather than singly, in the adaptive function of the brain \cite{Harris05}.  MVGC is well suited to disclosing causal relationships among these ensembles as a window onto underlying principles of brain operation.

Of course, the application of MVGC is not limited to neuroscience.  Multivariate interactions are likely to be important in a very broad range of application areas.  For example, genetic, metabolic, and transcriptional regulatory networks may be usefully decomposed into multivariate ensembles influencing other such ensembles \cite{ladroue:2009}.  Indeed, multivariate interactions may be important in any system, natural or artificial, which can be described in terms of multiple simultaneously acquired time series.

\subsection{Generalized variance vs total variance}

A different approach to multivariate causal analysis was recently proposed by Ladroue and colleagues \cite{ladroue:2009}.  This involved a measure (which we call trvMVGC) based on the trace of the residual covariance matrix (the total variance), rather than the determinant (the generalized variance).  Geweke \cite{Geweke82} provided the original justifications for the determinant form, but did not explicitly discuss the trace form.  As noted in Section 3 of Ref.~\cite{Geweke82}, Geweke's motivations included (i) MVGC is invariant under (linear) transformations of variables, and (ii) the maximum likelihood estimator of MVGC is asymptotically $\chi^2$-distributed for large samples; (there is no standard test statistic for trvMVGC).  In this paper we have substantially enhanced this list, in each case comparing MVGC explicitly with trvMVGC.  In summary: (iii) MVGC is fully equivalent to transfer entropy under Gaussian
assumptions, whereas for trvMVGC this equivalence only holds for the
univariate case; (iv) MVGC is invariant under \emph{all} (non-singular) linear transformations of the predictee variable, while trvMVGC is invariant only under conformal linear transformations (see below); (v) only MVGC is expandable as a sum of univariate G-causalities; (vi) MVGC but not trvMVGC admits a satisfactory spectral decomposition, inasmuch as it guarantees a consistent relationship with the corresponding time-domain formulation; (vii) only MVGC depends
on residual correlations, and through these accommodates in a natural way the influence of exogenous or latent variables, and (viii) the partial version of MVGC, pMVGC is decomposable in terms of non-partial MVGCs, but this is not true in general for trvMVGC.

All the above factors suggest that MVGC should be preferred to trvMVGC.  Taken individually they may differ in their significance but taken together they emphasize that MVGC, but not trvMVGC, provides a comprehensive and \emph{theoretically consistent} extension of standard G-causality to the multivariate case.  While this consistency is the most important reason to prefer MVGC to trvMVGC, let us consider further three of the individual properties. First, the equivalence with transfer entropy is important because it justifies the use of linear modeling for multivariate causal analysis, at least where Gaussian assumptions are reasonable. Second, the broader range of invariance is important because it means that MVGC is robust to a wider range of common inaccuracies during data collection, in particular those in which univariate variables are contaminated by contributions from other variables and in which different components of multivariate ensembles are differently scaled by measurement constraints. It is likely that this additional robustness will have significant practical importance in many experimental applications, for example in EEG and MEG where individual sensors detect signals from multiple neural sources and may differentially amplify these sources according to their distance from the sensors and their alignment with the cortical surface.  Finally, the lack of a satisfactory spectral version of trvMVGC, which we establish both theoretically and numerically (Section \ref{sec:sdecomp} and Figures \ref{fig:speccomp_hist} and \ref{fig:speccomp_single}), implies that frequency-domain results obtained using trvMVGC are unreliable, both in their magnitude and in their spectral profile.

Ladroue \emph{et al.} \cite{ladroue:2009} note Geweke's form (i.e.~MVGC) and suggest trvMVGC is preferable in view of possible numerical instabilities attending the computation of determinants for high-dimensional data.  However the existence of an expansion of MVGC in terms of univariate G-causalities \eqref{eq:Fexp} seems to counter this claim, since the univariate causalities would not be expected to be unstable.  Numerical simulations (Section \ref{sec:stability} and Figure \ref{fig:stability1}) confirm our view.

\subsection{Quantities derived from MVGC}

In the second part of the paper we used MVGC to derive several novel measures that have the potential to shed substantial new light on complex system dynamics.

First, MVGC leads immediately to a series of redefinitions of our previous ``causal density'' measure \cite{Seth05}, which aims to capture the dynamical complexity of a system's dynamics in terms of coexisting integration and differentiation.  Extension to the multivariate case allows causal density to be evaluated at multiple levels of description thus furnishing a more principled measure of dynamical complexity.  Causal density has been suggested as a measure of neural dynamics that captures certain aspects of consciousness \cite{SethEtal06}. It has been shown\footnote{In approximation.} to increase in response to perceived stimuli as compared to non-perceived stimuli in a visual masking task \cite{gaillard:2009}, and it captures the complex dynamics of small-world networks more effectively than does a prominent competing measure, neural complexity \cite{shanahan:2008}.  Multivariate causal density has the potential to further strengthen and generalize these contributions.

Second, MVGC can be used to generalize the concept of G-autonomy, which operationalizes the notion of autonomy as ``self causation'' \cite{seth:alife:2009}.  Multivariate G-autonomy is a significant enhancement because it deals with the case in which a group of variables may be jointly autonomous even though, individually, no variable is autonomous.   Our results therefore pave the way to informative application of this measure to complex systems.

Third, MVGC can be helpful in considering relations between microscopic and macroscopic levels of description of a system.  One approach is to consider how \emph{causally independent} a macroscopic variable is, with respect to its set of constituent micro-variables.  We have suggested that this notion can be used to identify parsimonious macro-variables by maximizing causal independence over a space of functions relating micro- and macro-variables.  Alternatively, the concept of G-emergence operationalizes the idea that an emergent macro-variable is both \emph{autonomous from} and \emph{causally dependent} on its underlying micro-level constituents.  Unlike the ``causal independence'' view, G-emergence may be better suited to characterizing the degree of emergence as opposed to identifying prospective macro-variables; G-emergence also explicitly measures micro-to-macro causal dependence rather than assuming that it is present.

Finally, the concepts of redundancy and synergy amongst variables have been recently introduced, via the use of a variant of the trvMVGC measure \cite{angelini}. These quantities aim at detecting functionally relevant partitions of a system by grouping variables according to their summed causal influences. Because of the advantages of MVGC over trvMVGC, we suggest it may be useful to redefine redundancy and synergy in terms of MVGC.

\subsection{Summary}

Models of complex systems typically contain large numbers of variables. Having a measure for directed interactions between groups of variables, as opposed to just single variables, provides a useful tool for the analysis of such systems. We have demonstrated that MVGC is such a measure, and we have provided a series of justifications, theoretical and numerical, to prefer it over a related measure, trvMVGC. Like all measures of directed interaction based on G-causality, MVGC can be measured for freely collected data, without perturbing or providing inputs to the system. Finally, in contrast to alternative approaches such as structural equation modeling \cite{kline:2005} or dynamic causal modeling \cite{Friston03}, MVGC can be applied with very little prior knowledge of the system under consideration.

\section*{Acknowledgements}
AKS is supported by EPSRC Leadership Fellowship EP/G007543/1, which also supports the work of ABB.  Support is also gratefully acknowledged from the Dr. Mortimer and Theresa Sackler Foundation.

\appendix

\section*{Appendix}

\section{Minimizing the determinant of the residuals covariance matrix} \label{apx:mindet}

We wish to show that minimizing the determinant $\dett{\Covs\beps}$, where $\beps =  \bX - A \cdot \bY$ as specified in \eqref{eq:linreg}, leads to the same values \eqref{eq:regcoef} for the regression coefficients $A$. We thus solve for $A$ in the simultaneous equations
\begin{equation}
    \pdiv{\dett{\Covs\beps}}{A_{i\alpha}} = 0\,,
\end{equation}
where $i$ runs from $1 \ldots n$, $\alpha$ from $1 \ldots m$ and $\Covs\beps$ is given by
\begin{equation}
    \Covs\beps = \Covs\bX - \Cov\bX\bY \trans A - A \trans{\Cov\bX\bY} + A \Covs\bY \trans A \label{eq:epsval} \,.
\end{equation}
We use the formula for an invertible square matrix $B$
\begin{equation}
    \pdiv{\dett B}{B_{jk}} = \dett B \bracr{B^{-1}}_{kj} \label{eq:dpdiv} \,.
\end{equation}
Assuming $\Covs\beps$ invertible and setting $W \equiv \dett{\Covs\beps} {\Covs\beps}^{-1}$ we have
\begin{align*}
    & \pdiv{\dett{\Covs\beps}}{A_{i\alpha}} = \sum_{j,k} \pdiv{\dett{\Covs\beps}}{\Covs\beps_{jk}} \pdiv{\Covs\beps_{jk}}{A_{i\alpha}} \\
    &= \sum_{j,k} W_{kj} \pdiv{\Covs\beps_{jk}}{A_{i\alpha}} \qquad\textrm{from \eqref{eq:dpdiv}} \\
    &= \sum_{j,k} W_{kj} \pdivf{A_{i\alpha}} \bracs{\Covs\bX - \Cov\bX\bY \trans A - A \trans{\Cov\bX\bY} + A \Covs\bY \trans A}_{jk} \qquad\textrm{from \eqref{eq:epsval}} \\
    &= \sum_{j,k} W_{kj} \pdivf{A_{i\alpha}} \bracs{- \sum_\beta \Cov\bX\bY_{j\beta} A_{k\beta} - \sum_\beta \Cov\bX\bY_{k\beta} A_{j\beta} + \sum_{\beta,\gamma} \Covs\bY_{\beta \gamma} A_{j\beta} A_{k\gamma}} \\
    &= \sum_{j,k} W_{kj} \bracs{- \sum_\beta \Cov\bX\bY_{j\beta} \delta_{ik}\delta_{\alpha\beta} - \sum_\beta \Cov\bX\bY_{k\beta} \delta_{ij}\delta_{\alpha\beta} + \sum_{\beta,\gamma} \Covs\bY_{\beta \gamma} \bracr{A_{j\beta} \delta_{ik}\delta_{\alpha\gamma} + A_{k\gamma} \delta_{ij}\delta_{\alpha\beta}}} \\
	&= - \sum_j W_{ij} \Cov\bX\bY_{j\alpha} - \sum_k  W_{ki} \Cov\bX\bY_{k\alpha} + \sum_{\beta,j}  W_{ij} \Covs\bY_{\beta \alpha} A_{j\beta} + \sum_{\gamma,k}  W_{ki} \Covs\bY_{\alpha\gamma} A_{k\gamma} \\
    &= 2 \bracc{W \bracs{A \Covs\bY - \Cov\bX\bY}}_{i\alpha}
\end{align*}
after gathering terms and simplifying, and Eq.~\eqref{eq:regcoef} follows.

\section{Proof of expansion of multivariate Granger causality} \label{expandproof}
Here we prove Eq.~\eqref{eq:Fexp}. We consider the case of there being no conditional third variable, since the extension to this case is trivial. We first expand in terms of predictor variables according to
\begin{align}
&\gc\bY\bX = \logt{\frac{\dett{\Covc\bX\blX}} {\dett{\Covc\bX{\blX\mc\blY}}}} \nonumber \\
&= \logt{\frac{\dett{\Covc\bX\blX} \cdot \dett{\Covc\bX{\blX\mc\blY_1}} \cdot \dett{\Covc\bX{\blX\mc\blY_1\mc\blY_2}} \cdots
 \dett{\Covc\bX{\blX\mc\blY_1\mc\ldots\blY_{m-1}}}} {\dett{\Covc\bX{\blX\mc\blY_1}} \cdot \dett{\Covc\bX{\blX\mc\blY_1\mc\blY_2}} \cdots \dett{\Covc\bX{\blX\mc\blY_1\mc\ldots\blY_m}}}} \nonumber \\
 &= \logt{\frac{\dett{\Covc\bX\blX}} {\dett{\Covc\bX{\blX\mc\blY_1}}}} + \logt{\frac{\dett{\Covc\bX{\blX\mc\blY_1}}} {\dett{\Covc\bX{\blX\mc\blY_1\mc\blY_2}}}} + \cdots \nonumber \\
  & \phantom= + \logt{\frac{\dett{\Covc\bX{\blX\mc\blY_1\mc\ldots\blY_{m-1}}}} {\dett{\Covc\bX{\blX\mc\blY_1\mc\ldots\blY_m}} }} \nonumber \\
& = \gc{Y_1}\bX + \cgc{Y_2}\bX{Y_1}+\cgc{Y_3}\bX{Y_1\mc Y_2}+\cdots + \cgc{Y_m}\bX{Y_1\mc Y_2\mc\ldots\mc Y_{m-1} }\,.\label{fprime1}
\end{align}
To expand in terms of predictees we use the expansion
\begin{equation}
\dett{\Covc\bX\bW} = \Covs{X_1}\Covc{X_2}{\bW\mc X_1}\Covc{X_3}{\bW\mc X_1\mc X_2}\cdots \Covc{X_n}{\bW\mc X_1\mc\ldots X_{n-1}}\,,
\end{equation}
which follows from repeated application of Eq.~\eqref{eq:cxyid}. We obtain
\begin{align}
& \gc{Y_1}\bX = \logt{\frac{\dett{\Covc\bX\blX}} {\dett{\Covc\bX{\blX\mc\blY_1}}}} \nonumber \\
&= \logt{\frac{\Covc{X_1}\blX \Covc{X_2}{\blX\mc X_1} \cdots \Covc{X_n}{\blX\mc X_1\mc X_2\mc \ldots \mc X_{n-1}}} {\Covc{X_1}{\blX\mc\blY_1} \Covc{X_2}{\blX\mc\blY_1\mc X_1} \cdots \Covc{X_n}{\blX\mc\blY_1\mc X_1\mc X_2\mc \ldots \mc X_{n-1}}}} \nonumber \\
&= \cgc{Y_1}{X_1}\bX + \cgc{Y_1}{X_2}{\bX\mc X_1^0} + \cgc{Y_1}{X_3}{\bX\mc X_1^0\mc X_2^0} + \cdots + \cgc{Y_1}{X_n}{\bX\mc X_1^0\mc X_2^0\mc\ldots\mc X_{n-1}^0}\,,
\end{align}
and similar for the other components of the sum in Eq.~\eqref{fprime1}, from which the result follows.

\section{Partial covariance of residuals for two variables jointly dependent on a third} \label{apx:pgc}

Given the regressions
\begin{align}
	\bX & = A \cdot \bW + \beps\,, \nonumber\\
	\bZ & = B \cdot \bW + \beeta\,,
\end{align}
where the regression coefficients $A,B$ are derived from an ordinary least squares, Yule-Walker or equivalent procedure, we show that
\begin{equation}
	\Covc\beps\beeta = \Covc\bX{\bZ\mc\bW}\,, \label{eq:rcepet}
\end{equation}
assuming that all (partial) covariance matrices which appear below are invertible.
We have
\begin{align}
	\Covs\beps & = \Covc\bX\bW\,, \nonumber\\
	\Covs\beeta & = \Covc\bZ\bW\,, \\
	\Cov\beps\beeta & = \Covc{\bX,\bZ}\bW\,. \nonumber
\end{align}
Thus we may calculate that
\begin{equation}
	\Covc\beps\beeta = \Covc\bX\bW - \Covc{\bX,\bZ}\bW \Covc\bZ\bW^{-1} \Covc{\bZ,\bX}\bW\,. \label{eq:epet}
\end{equation}
Using the block matrix inversion formula for $\Covs{\bZ\mc\bW}$,  we may also calculate that
\begin{equation}
	\Covc\bX{\bZ\mc\bW} = \Covs\bX  - \Covc{\bX,\bZ}\bW \Covc\bZ\bW^{-1} \Cov\bZ\bX - \Covc{\bX,\bW}\bZ \Covc\bW\bZ^{-1} \Cov\bW\bX\,. \label{eq:epet1}
\end{equation}
Now expanding the $\Covc\bX\bW \equiv \Covs\bX - \Cov\bX\bW\Covs\bW^{-1}\Cov\bW\bX$ term in \eqref{eq:epet}, we find using  \eqref{eq:epet1} that \eqref{eq:rcepet} is equivalent to
\begin{multline*}
	\Cov\bX\bW\Covs\bW^{-1}\Cov\bW\bX + \Covc{\bX,\bZ}\bW \Covc\bZ\bW^{-1} \Covc{\bZ,\bX}\bW \\
	= \Covc{\bX,\bZ}\bW \Covc\bZ\bW^{-1} \Cov\bZ\bX + \Covc{\bX,\bW}\bZ \Covc\bW\bZ^{-1} \Cov\bW\bX\,.
\end{multline*}
Or, rearranging and factorizing,
\begin{multline}
	\bracs{\Cov\bX\bW\Covs\bW^{-1} - \Covc{\bX,\bW}\bZ \Covc\bW\bZ^{-1}} \Cov\bW\bX \\
	= \Covc{\bX,\bZ}\bW \Covc\bZ\bW^{-1} \bracs{\Cov\bZ\bX - \Covc{\bZ,\bX}\bW}\,. \label{eq:epet3}
\end{multline}
Now the term in square brackets on the RHS of \eqref{eq:epet3} simplifies to $\Cov\bZ\bW\Covs\bW^{-1}\Cov\bW\bX$ so that, factoring out $\Cov\bW\bX$, \eqref{eq:epet3} is equivalent to
\begin{multline}
	\bracs{\Cov\bX\bW\Covs\bW^{-1} - \Covc{\bX,\bW}\bZ \Covc\bW\bZ^{-1}
	- \Covc{\bX,\bZ}\bW \Covc\bZ\bW^{-1} \Cov\bZ\bW\Covs\bW^{-1}} \\
	 \times \Cov\bW\bX = 0\,. \label{eq:epet4}
\end{multline}
We now show that the term in the square brackets in \eqref{eq:epet4} is zero; \ie\ that
\begin{equation}
	\Cov\bX\bW\Covs\bW^{-1} - \Covc{\bX,\bW}\bZ \Covc\bW\bZ^{-1}
	- \Covc{\bX,\bZ}\bW \Covc\bZ\bW^{-1} \Cov\bZ\bW\Covs\bW^{-1} = 0\,, \label{eq:epet5}
\end{equation}
thus proving \eqref{eq:rcepet}. Rearranging and factoring out $\Covs\bW^{-1}$, \eqref{eq:epet5} becomes
\begin{equation*}
	\bracs{\Cov\bX\bW - \Covc{\bX,\bZ}\bW \Covc\bZ\bW^{-1} \Cov\bZ\bW}\Covs\bW^{-1}
	= \Covc{\bX,\bW}\bZ \Covc\bW\bZ^{-1}\,,
\end{equation*}
or, multiplying through on the right by $\Covc\bW\bZ$,
\begin{equation*}
	\bracs{\Cov\bX\bW - \Covc{\bX,\bZ}\bW \Covc\bZ\bW^{-1} \Cov\bZ\bW}\Covs\bW^{-1} \Covc\bW\bZ
	= \Covc{\bX,\bW}\bZ\,.
\end{equation*}
Expanding $\Covc\bW\bZ$, factorizing and rearranging again, we get
\begin{multline*}
	\bracs{\Cov\bX\bZ - \Cov\bX\bW \Covs\bW^{-1} \Cov\bW\bZ} \Covs\bZ^{-1} \Cov\bZ\bW \\
	= \Covc{\bX,\bZ}\bW \Covc\bZ\bW^{-1}\Cov\bZ\bW\Covs\bW^{-1}\Covc\bW\bZ\,,
\end{multline*}
or, since the term in square brackets on the LHS is just $\Covc{\bX,\bZ}\bW$,
\begin{equation*}
	\Covc{\bX,\bZ}\bW \bracs{\Covs\bZ^{-1} \Cov\bZ\bW- \Covc\bZ\bW^{-1}\Cov\bZ\bW\Covs\bW^{-1}\Covc\bW\bZ}\,.
\end{equation*}
We now show that, again, the term in square brackets is zero; \ie\ that
\begin{equation}
	\Covs\bZ^{-1} \Cov\bZ\bW = \Covc\bZ\bW^{-1}\Cov\bZ\bW\Covs\bW^{-1}\Covc\bW\bZ\,. \label{eq:epet6}
\end{equation}
Multiplying through on the left by $\Covc\bZ\bW$, \eqref{eq:epet6} is equivalent to
\begin{equation*}
	\Covc\bZ\bW\Covs\bZ^{-1} \Cov\bZ\bW = \Cov\bZ\bW\Covs\bW^{-1}\Covc\bW\bZ\,,
\end{equation*}
which follows immediately on expanding $\Covc\bZ\bW$ and $\Covc\bW\bZ$, thus establishing \eqref{eq:rcepet}.


\begin{thebibliography}{10}

\bibitem{DingEtal06}
M.~Ding, Y.~Chen and S.~Bressler.
\newblock Granger causality: {B}asic theory and application to neuroscience.
\newblock In S.~Schelter, M.~Winterhalder, and J.~Timmer, editors, {\em
  Handbook of Time Series Analysis}, 438--460. Wiley, Wienheim, 2006.

\bibitem{Friston03}
K.~Friston, L.~Harrison and W.~Penny.
\newblock Dynamic causal modeling.
\newblock {\em Neuroimage}, 19(4):1273--302, 2003.

\bibitem{Schreiber00}
T.~Schreiber.
\newblock Measuring information transfer.
\newblock {\em Phys Rev Lett}, 85(2):461--4, 2000.

\bibitem{ladroue:2009}
C.~Ladroue, S.~Guo, K.~Kendrick and J.~Feng.
\newblock {Beyond} element-wise interactions: Identifying complex interactions
  in biological processes.
\newblock {\em PLoS One}, 4:e6899--e6899, 2009.

\bibitem{Geweke82}
J.~Geweke.
\newblock Measurement of linear dependence and feedback between multiple time
  series.
\newblock {\em J Am Stat Assoc}, 77(378):304--313, 1982.

\bibitem{zhou:2009}
Z.~Zhou, Y.~Chen, M.~Ding, P.~Wright, Z.~Lu, and Y.~Liu.
\newblock {Analyzing} brain networks with {PCA} and conditional {Granger}
  causality.
\newblock {\em Hum Brain Mapp}, 30:2197--2206, 2009.

\bibitem{wiener:1956}
N.~Wiener.
\newblock The theory of prediction.
\newblock In E.~F.~Beckenbach, editor, {\em Modern Mathematics for
  Engineers}. McGraw Hill, New York, NY, 1956.
  
\bibitem{Granger69}
C.~W.~J.~Granger.
\newblock Investigating causal relations by econometric models and
  cross-spectral methods.
\newblock {\em Econometrica}, 37:424--438, 1969.
  
\bibitem{bressler:2010}
S.~L.~Bressler and A.~K.~Seth.
\newblock Wiener-Granger causality: {A} well established methodology.
\newblock {\em Neuroimage}, x:xx--xx, 2010.

\bibitem{seth:2009a}
A.~K.~Seth.
\newblock Explanatory correlates of consciousness: {T}heoretical and
  computational challenges.
\newblock {\em Cognitive Computation}, 1(1):50--63, 2009.

\bibitem{seth:alife:2009}
A.~K.~Seth.
\newblock Measuring autonomy and emergence via Granger causality.
\newblock {\em Artificial Life}, 16(2), 2009.

\bibitem{Kendall79}
M.~G.~Kendall and A.~Stuart.
\newblock {\em The advanced theory of statistics, volume 2: ``Inference and
  Relationship''}.
\newblock Griffin, London, 1979.

\bibitem{Barnett:2009a}
L.~Barnett, A.~B.~Barrett and A.~K.~Seth.
\newblock Granger causality and transfer entropy are equivalent for {G}aussian
  variables.
\newblock {\em Phys Rev Lett}, 103:238701, 2009.

\bibitem{Granger63}
C.~W.~J.~Granger.
\newblock Economic processes involving feedback.
\newblock {\em Inform Control}, 6:28--48, 1963.

\bibitem{Whittle53}
P.~Whittle.
\newblock The analysis of multiple stationary time series.
\newblock {\em J Royal Stat Soc B}, 15(1):125--139, 1953.

\bibitem{Wald43}
A.~Wald.
\newblock Tests of statistical hypotheses concerning several parameters when
  the number of observations is large.
\newblock {\em T Am Math Soc}, 54(3):426--482, 1943.

\bibitem{Seth:2010}
A.~K.~Seth.
\newblock A MATLAB toolbox for Granger causal connectivity analysis.
\newblock {\em J Neurosci Meth}, 186:262--273, 2010.

\bibitem{Geweke84}
J.~Geweke.
\newblock Measures of conditional linear dependence and feedback between time
  series.
\newblock {\em J Am Stat Assoc}, 79(388):907--915, 1984.

\bibitem{KaiserSchreiber02}
A.~Kaiser and T.~Schreiber.
\newblock Information transfer in continuous processes.
\newblock {\em Physica D}, 166:43--62, 2002.

\bibitem{ChenEtal04}
Y.~Chen, G.~Rangarajan, J.~Feng and M.~Ding.
\newblock Analyzing multiple nonlinear time series with extended {G}ranger
  causality.
\newblock {\em Phys Lett A}, 324:26--35, 2004.

\bibitem{Marinazzo08}
D.~Marinazzo, M.~Pellicoro, and S.~Stramaglia.
\newblock Kernel method for nonlinear Granger causality.
\newblock {\em Phys Rev Lett}, 100(14):144103, 2008.

\bibitem{Seth05}
A.~K.~Seth.
\newblock Causal connectivity of evolved neural networks during behavior.
\newblock {\em Network: Computation in Neural Systems}, 16:35--54, 2005.

\bibitem{SpornsEtal00}
O.~Sporns, G.~Tononi and G.M. Edelman.
\newblock Theoretical neuroanatomy: {R}elating anatomical and functional
  connectivity in graphs and cortical connection matrices.
\newblock {\em Cerebral Cortex}, 10:127--141, 2000.

\bibitem{Chen06}
Y.~Chen, S.~L.~Bressler and M.~Ding.
\newblock Frequency decomposition of conditional {G}ranger causality and
  application to multivariate neural field potential data.
\newblock {\em J Neurosci Meth}, 150:228--237, 2006.

\bibitem{Rozanov67}
Y.~A.~Rozanov.
\newblock {\em Stationary Random Processes}.
\newblock Holden-Day, San Francisco, CA, 1967.

\bibitem{Doob53}
J.~Doob.
\newblock {\em Stochastic Processes}.
\newblock John Wiley, New York, NY, 1953.

\bibitem{Akaike74}
H.~Akaike.
\newblock A new look at the statistical model identification.
\newblock {\em IEEE Trans Autom Control}, 19:716-723, 1974.

\bibitem{Schwartz78}
G.~Schwartz.
\newblock Estimating the dimension of a model.
\newblock {\em The Annals of Statistics}, 5(2):461-464, 1978.

\bibitem{guoetal:2008}
S.~Guo, A.~K.~Seth, K.~Kendrick, C.~Zhou and J.~Feng.
\newblock Partial Granger causality: Eliminating exogenous inputs and latent
  variables.
\newblock {\em J Neurosci Meth}, 172:79--93, 2008.

\bibitem{baccala:2001}
L.~A.~Baccal{\'{a}} and K.~Sameshima.
\newblock {Partial} directed coherence: a new concept in neural structure
  determination.
\newblock {\em Biol Cybern}, 84:463--474, 2001.

\bibitem{sporns:2007}
O.~Sporns.
\newblock Complexity.
\newblock {\em Scholarpedia}, 2(10):1623, 2007.

\bibitem{SethEtal06}
A.~K.~Seth, E.~Izhikevich, G.~N.~Reeke and G.~M.~Edelman.
\newblock Theories and measures of consciousness: {A}n extended framework.
\newblock {\em P Natl Acad Sci, USA},
  103(28):10799--10804, 2006.

\bibitem{shanahan:2008}
M.~Shanahan.
\newblock {Dynamical} complexity in small-world networks of spiking neurons.
\newblock {\em Phys Rev E Stat Nonlin Soft Matter Phys}, 78:041924--041924,
  2008.

\bibitem{TononiEtal94}
G.~Tononi, O.~Sporns and G.~M.~Edelman.
\newblock A measure for brain complexity: {R}elating functional segregation and
  integration in the nervous system.
\newblock {\em P Natl Acad Sci, USA},
  91:5033--5037, 1994.
  
\bibitem{Bedau97}
M.~A.~Bedau.
\newblock Weak emergence.
\newblock {\em Philosophical Perspectives}, 11:375--399, 1997.
  
\bibitem{ShaliziMoore06}
C.~R.~Shalizi and C.~Moore.
\newblock What is a macrostate: {S}ubjective observations and objective
  dynamics.
\newblock http://arxiv.org/abs/cond-mat/0303625, 2003.

\bibitem{shalizi:2006}
C.~R.~Shalizi, R.~Haslinger, J-B.~Rouquier, K.~L.~Klinkner and C.~Moore.
\newblock {Automatic} filters for the detection of coherent structure in
  spatiotemporal systems.
\newblock {\em Phys Rev E Stat Nonlin Soft Matter Phys}, 73:036104--036104,
  2006.

\bibitem{Seth07a}
A.~K.~Seth.
\newblock Granger causality.
\newblock {\em Scholarpedia}, 2(7):1667, 2007.

\bibitem{Hebb49}
D.~O.~Hebb.
\newblock {\em The organization of behavior}.
\newblock Wiley, New York, NY, 1949.

\bibitem{Harris05}
K.~D.~Harris.
\newblock Neural signatures of cell assembly organization.
\newblock {\em Nat Rev Neurosci}, 6:399--407, 2005.

\bibitem{gaillard:2009}
R.~Gaillard, S.~Dehaene, C.~Adam, S.~Cl{\'{e}}menceau, D.~Hasboun, M.~Baulac, L.~Cohen and L.~Naccache.
\newblock {Converging} intracranial markers of conscious access.
\newblock {\em PLoS Biol}, 7:e61--e61, 2009.


\bibitem{angelini}
L.~Angelini, M.~de Tommaso, D.~Marinazzo, L.~Nitti, M.~Pellicoro and S.~Stramaglia.
\newblock Redundant variables and Granger causality.
\newblock {\em Phys Rev E}, in press, 2010.

\bibitem{kline:2005}
R.~B.~Kline.
\newblock {\em Principles and practice of structural equation modeling}.
\newblock Guilford Press, New York, NY, 2005.

\end{thebibliography}

\end{document}